\documentclass[prb,aps,superscriptaddress,twocolumn,nofootinbib]{revtex4-1}

\usepackage{mathrsfs}
\usepackage{amsfonts}
\usepackage{amssymb}
\usepackage{amsmath}
\usepackage{graphicx}
\usepackage[usenames,dvipsnames]{color}
\usepackage[colorlinks=true,citecolor=blue,linkcolor=magenta]{hyperref}
\usepackage{lmodern}
\usepackage{amsthm}
\usepackage{dsfont,natbib,caption,subcaption}
\usepackage{textcomp}
\usepackage{tikz}
\usepackage{tikz-3dplot}
\usetikzlibrary{shapes,arrows}
\usepackage{cancel}
\usepackage{makecell}
\usepackage{soul}

\AtBeginDocument{%
    \newwrite\bibnotes
    \def\bibnotesext{Notes.bib}
    \immediate\openout\bibnotes=\jobname\bibnotesext
    \immediate\write\bibnotes{@CONTROL{REVTEX41Control}}
    \immediate\write\bibnotes{@CONTROL{%
    apsrev41Control,author="08",editor="1",pages="1",title="1",year="1"}}
     \if@filesw
     \immediate\write\@auxout{\string\citation{apsrev41Control}}%
    \fi
}%

\newcommand{\be}{\begin{equation}}
\newcommand{\ee}{\end{equation}}
\newcommand{\ket}[1]{\vert{ #1 }\rangle}
\newcommand{\bra}[1]{\langle{ #1 }\vert}

\newcommand{\braket}[2]{\langle #1 \vert #2 \rangle}
\newcommand{\pt}{\partial}
\newcommand{\abs}[1]{\vert#1\vert}

\newcommand{\zee}{\mathbb{Z}}
\newcommand{\ztwo}{\mathbb{Z}_2}
\newcommand{\kvec}{\mathbf{k}}

\newcommand{\TR}{\mathcal{T}}
\newcommand{\PH}{\mathcal{P}}
\newcommand{\CR}{\mathcal{R}}
\newcommand{\cc}{\mathcal{K}}
\newcommand{\sew}{\mathbb{S}}
\newcommand{\trans}{\mathrm{T}}
\newcommand{\iden}{\mathbb{I}}
\newcommand{\VB}{\mathcal{E}}

\begin{document}

\title{Topological invariants beyond symmetry indicators: Boundary diagnostics for twofold rotationally symmetric superconductors} 

\affiliation{C. N. Yang Institute for Theoretical Physics and Department of Physics and Astronomy, State University of New York at Stony Brook, Stony Brook, NY 11794, USA\\
$^2$Condensed Matter Theory Center and Joint Quantum Institute, University of Maryland, College Park, MD 20742, USA\\
$^3$Department of Physics, University of Notre Dame, Notre Dame, IN 46556, USA\\}

\author{Yanzhu Chen$^{1}$}
\altaffiliation[Present address: ]{Department of Physics, Virginia Tech, Blacksburg, VA 24061, USA}
\author{Sheng-Jie Huang$^{2}$}
\author{Yi-Ting Hsu$^{3}$}
\author{Tzu-Chieh Wei$^{1}$}

\date{\today}							

\begin{abstract}
Topological crystalline superconductors are known to have possible higher-order topology, which results in Majorana modes on $d-2$ or lower-dimensional boundaries. Given the rich possibilities of boundary signatures, it is desirable to have topological invariants that can predict the type of Majorana modes from band structures. Although symmetry indicators, a type of invariant that depends only on the band data at high-symmetry points, have been proposed for certain crystalline superconductors, there exist symmetry classes in which symmetry indicators fail to distinguish superconductors with different Majorana boundaries. Here, we systematically obtain topological invariants for an example of this kind, two-dimensional time-reversal symmetric superconductors with twofold rotational symmetry $C_2$. First, we show that the nontrivial topology is independent of band data on the high-symmetry points by conducting a momentum-space classification study. Then, from the resulting K groups, we derive calculable expressions for four $\mathbb{Z}_2$ invariants defined on high-symmetry lines or general points in the Brillouin zone. Finally, together with a real-space classification study, we establish the bulk-boundary correspondence and show that the four $\mathbb{Z}_2$ invariants can predict Majorana boundary types from band structures. Our proposed invariants can fuel practical material searches for $C_2$-symmetric topological superconductors featuring Majorana edge and corner modes. 
\end{abstract}

\maketitle

\section{Introduction}

The classification of non-interacting fermionic topological phases of matter has been extensively studied and is complete for systems with the basic internal symmetries concerning time-reversal, particle-hole, and chiral symmetries~\cite{Schnyder2008,Kitaev2009,Schnyder2009,Ryu2010,Stone2010,Hsieh2012,Ryu2012,Abramovici2012,Freed2013,Kennedy2015,Ludwig2015,Chiu2016}. 
Due to the bulk-boundary correspondence, it is well-known that internal symmetries can protect gapless modes on $(d-1)$-dimensional boundaries in a $d$-dimensional topological phase.
In systems with crystalline symmetries, however, it has been realized more recently that the presence of these spatial symmetries enriches the classification of the bulk topology~\cite{Fu2011,Hughes2011,Teo2013,Shiozaki2014,Isobe2015,Shiozaki2017,Khalaf2018_1} and induces new phases with gapless modes in ($d-2$)- or even lower dimensional boundaries.  
Insulators and superconductors with such boundary signatures are dubbed higher-order topological phases, which have attracted great theoretical and experimental interest in the past few years~\cite{Benalcazar2017, Benalcazar2017multipole,Langbehn2017,Song2017rot,Schindler2018,Neupert2018book,Shapourian2018,
Guido2018,Geier2018,Trifunovic2019,Zhang2019,Hsu2020,Zhang2020,Vu2020,Tiwari2020,Wu2021,Huang2021,Jahin2021,Chew2021,Li2021_1}. For example, instead of Majorana edge modes, a two-dimensional (2D) higher-order topological superconductor (TSC) under the protection of the inversion symmetry can host Majorana zero modes at opposite corners~\cite{Hsu2020,Huang2021}. 

Topological invariants that can predict the type of boundary modes detectable in experiments for a given topological crystalline phase are therefore desirable. 
For the insulating phases, it has been shown that both the bulk topology and boundary signatures can sometimes be characterized in terms of the crystalline symmetry eigenvalues of occupied bands on the high symmetry points in the Brillouin zone (BZ)~\cite{Fu2007_1,Fu2010,Sato2010,Fang2012}.
This type of invariants, which are calculated by band structure information on the high-symmetry points only, are generally dubbed \textit{symmetry indicators}~\cite{Slager2013,Kruthoff2017,Po2017,Bradlyn2017,Khalaf2018,Ono2018,Shiozaki2019,Po2020,Huang2021}. 
For superconducting phases, symmetry indicators based on the crystalline symmetry eigenvalues of occupied Bogoliubov-de Gennes (BdG) bands at high-symmetry points have also been used to distinguish different types of bulk topology as well as Majorana boundary modes~\cite{Shiozaki2019,Ono2019,Hsu2020,Ono2020,Skurativska2020,Geier2020,Ono2021,Huang2021}.

Despite being powerful and easy to compute, symmetry indicators cannot distinguish all crystalline topological phases. For insulators of this kind, proper invariants have been proposed based on the band topology in higher-dimensional subspaces rather than high-symmetry points of the BZ~\cite{Schindler2019,Bouhon2020,Cano2021}. However, much less is known for topological crystalline superconductors that cannot be characterized by the symmetry indicators. A prior work~\cite{Vu2020} have found lattice models for two-fold rotational ($C_2$) and time-reversal symmetric superconductors where symmetry indicators failed to distinguish different topological phases. For such classes of topological crystalline superconductors, a general and systematic scheme for obtaining momentum-space topological invariants that are capable of fully characterizing the bulk topology as well as the boundary signatures remains absent. 

Our goal is to study the classification of topological crystalline superconductors that \textit{cannot} be characterized by symmetry indicators and to systematically derive topological invariants capable of fully characterizing both the bulk topology and the Majorana boundary signatures. 
In this work, we focus on a case study of 2D time-reversal superconductors with $C_2$ rotational symmetry, motivated by the puzzling failure of symmetry indicators previously observed in lattice models of this symmetry class~\cite{Vu2020}.  
In fact, by computing the classification group in the momentum space for our case study, we find that the topology is indeed trivial on the high-symmetry points so that symmetry indicators are not the proper topological invariants. Instead, we find nontrivial classification groups of $\ztwo$ on high-symmetry lines and on general points in the BZ. 
We therefore seek to derive practically calculable expressions for the corresponding $\ztwo$ topological invariants that not only fully characterize the bulk topology but can also diagnose the Majorana boundary type for 2D time-reversal and $C_2$-symmetric superconductors.

The main challenge in deriving such boundary diagnostics that take band information as inputs is to establish the bulk-boundary correspondence for the considered topological crystalline superconductors.
In this work, we approach this problem following a three-step scheme that some of us previously developed~\cite{Huang2021}. 
The first step is to study the classification in the momentum space by calculating the K group for the considered superconductors, and to derive calculable expressions for the momentum-space topological invariants accordingly.  
Given that the K group is generally difficult to compute in the presence of crystalline symmetries, we approximate the K group using a mathematical tool called Atiyah-Hirzebruch spectral sequence (AHSS)~\cite{Shiozaki2018,Stehouwer2018}. 
The key idea of the AHSS method is to decompose the full $C_2$-symmetric BZ into different subspaces with only effective internal symmetry such that one can directly compute the contribution from each of the subspaces to the total K group. 
The approximation is then improved order by order, which eventually converges and provides exact information about the K group.

Importantly, from the AHSS results, we not only find that the nontrivial topology is fully encoded in the 1D and 2D subspaces (i.e. the high-symmetry lines and general points in the BZ), but also obtain various constraints and hints for writing down explicit expressions for the corresponding $\ztwo$ topological invariants.  
In particular, the invariants we find are based on \textit{the sewing matrix} of $C_2T$ symmetry of the superconductors, where $T$ denotes the time-reversal symmetry. We further propose an alternative expression that can be practically computed for more general cases by identifying these $\ztwo$ invariants as the Stiefel-Whitney (S-W) classes of a real vector bundle defined by the occupied BdG states of the superconductors. Similar sewing-matrix-based invariants and the interpretation as the S-W classes have also been written down for topological crystalline insulators and semimetals~\cite{Ahn2018,Ahn2019,Bouhon2020}. The resulting four $\ztwo$ invariants $\{\nu_{1a},\nu_{1b},\nu_{1c},\nu_{2}\}$, where the first three and the last one depend on the BdG band data on the high-symmetry lines and general points in the BZ, respectively, together can fully discern 2D time-reversal and $C_2$-symmetric superconducting phases with distinct bulk topology. 

Next, we establish how these momentum-space invariants correspond to various Majorana boundary signatures in the real space. Specifically, in the second step of our protocol, we conduct a dual classification study in the real space using a method called \textit{topological crystal approach} \cite{Huang2017,Song2017,shiozaki2018generalized,Song2019,Song2020,Song2020real,Huang20204d,huang2021effective}.
This real-space analysis allows us to construct the representative state, learn the protecting symmetries, and obtain the Majorana boundary modes for each of the topologically distinct phases. 
For the protecting symmetries, we define weak and strong phases to be the phases protected by translational and other symmetries, respectively. For the boundary signatures, we define second-order and first-order superconductors to be the states with Majorana corner modes and 1D Majorana modes on all or certain-directional edges, respectively, regardless of the protecting symmetries. When a state supports more than one type of boundary modes, we define the order to be the lowest co-dimension of the boundary modes so that the order indicates the type of Majorana boundaries we expect to be experimentally detectable.  
Our result shows that all possible phases are given by  combinations of four fundamental phases, including two first-order weak phases with Majorana bands on certain edges, a first-order strong phase with helical Majorana edge modes, and a second-order strong phase with Majorana corner modes.  

Finally in the third step, equipped with the $\ztwo$ momentum-space topological invariants $\{\nu_{1a},\nu_{1b},\nu_{1c},\nu_{2}\}$ and all possible types of Majorana boundary modes, we establish the bulk-boundary correspondence by computing our invariants for the real-space representative states. 
Crucially, since these representative states are adiabatically connected to all other states in each of the phases, the boundary diagnostics that we aim to derive are capable of characterizing each of the phases. 
By matching the momentum-space and the real-space bases, we find that the final invariants $\{z_{1},z_{2},z_{3},z_{4}\}$ that can diagnose the type of Majorana boundary modes are given by certain combinations of the original invariants  $\{\nu_{1a},\nu_{1b},\nu_{1c},\nu_{2}\}$.  These practically calculable boundary diagnostics for 2D time-reversal and $C_2$-symmetric superconductors are our central results, as we summarize in Table~\ref{tab:new_inv}. 

The rest of the paper is organized as follows. In Sec.~\ref{sec:setup}, we specify the symmetry class of our case study and introduce key concepts about the equivariant K theory~\cite{Shiozaki2018,Stehouwer2018,Huang2021}. In Sec.~\ref{sec:AHSS}, we conduct the first step in our protocol. We first review the mathematical tool we use for the momentum-space classification, that is the AHSS. We then present the classification results for our case study and leave the details of the calculation in Appendix~\ref{app:AHSS}. In particular, we show that the proper topological invariants for our case study are \textit{not} symmetry indicators.  
In Sec.~\ref{sec:sew}, we derive practically calculable expressions for the momentum-space topological invariants from our classification results in the framework of the equivariant K theory. In particular, we propose expressions based on the homotopy classes of a sewing matrix as well as the S-W classes of the real vector bundles constructed from our system.
In Sec.~\ref{sec:real}, we first conduct the second step of our protocol and perform a real-space classification study using the topological crystal approach. From the analysis, we obtain the Majorana boundary signatures for all topologically distinct phases. 
We then perform the third step of our protocol and establish the bulk-boundary correspondence for our case study by explicitly computing our momentum-space topological invariants for each of the real-space representative states.  
In Sec.~\ref{sec:edge}, we explain how to construct the final Majorana boundary diagnostics from the momentum-space topological invariants. The results are presented in Table~\ref{tab:new_inv}. 
Finally in Sec.~\ref{sec:discussion}, we summarize our results and discuss several remarks.

\section{Case study}
\label{sec:setup}

In this section, we will first briefly introduce the equivariant K group, a mathematical object that gives the classifications of non-interacting fermionic systems with crystalline symmetries. We will then specify the symmetry group of the 2D superconductors for our case study in this work.

A gapped non-interacting fermionic system may possess one or more of the following internal symmetries: time reversal, particle-hole, and chiral symmetry. It then falls in one of the ten Altland-Zirnbauer (AZ) classes~\cite{Schnyder2008,Ryu2010,Ludwig2015,Chiu2016} describing whether each of the three symmetries is present and if so, whether it squares to $1$ or $-1$. In order to classify the phases of matter in such a system, one can adiabatically deform the Bloch Hamiltonian or the BdG Hamiltonian, while preserving the symmetries, so that the occupied and unoccupied bands are at two flat levels respectively. The problem of classifying the Hamiltonians is thus equivalent to classifying the occupied states treated as a vector bundle with the BZ as its base space. This allows the classification to be studied under the framework of the K theory~\cite{Horava2005,Kitaev2009,Wen2012,Chiu2016}. 

In addition to the symmetries in the AZ class, the presence of crystalline symmetries leads to finer separation of phases. A formalism called twisted equivariant K theory has been developed to accommodate crystalline symmetries in the classification of gapped non-interacting fermionic systems~\cite{Shiozaki2017,Shiozaki2018,Stehouwer2018}. In this formalism, different phases in the AZ class $n$ form an Abelian group $^\phi K_G^{(\tau,c),-n}(BZ)$, where the symmetry action of the group $G$ is further specified by $\phi$, $\tau$, and $c$. The group homomorphism $\phi:G\rightarrow\ztwo$ denotes whether an element $g\in G$ is unitary ($\phi(g)=1$) or anti-unitary ($\phi(g)=-1$). The factor system $\tau$ gives the $U(1)$ phase factor in the group multiplication
\be
	U_g (g^\prime \kvec) U_{g^\prime} (\kvec) = e^{i\tau_{g,g^\prime} (gg^\prime\kvec)} U_{gg^\prime} (\kvec),
\ee
where $\kvec$ under the action of $g$ is changed to $g\kvec$. A nontrivial $\tau$ arises from the non-symmorphic symmetries in $G$ and/or the projective representation of $G$. The group homomorphism $c:G\rightarrow\ztwo$ determines the commutation relation between $g\in G$ and the Hamiltonian $H$ by
\be
	U_g(\kvec) H(\kvec) U_g(\kvec)^{-1} = c(g) H(g\kvec). 
\ee
When the system lives on a $d$-dimensional lattice with translational symmetries, the BZ is a torus $T^d$. One can make use of the mathematical tool of AHSS to construct an approximation to $^\phi K_G^{(\tau,c),-n}(BZ)$~\cite{Shiozaki2017,Shiozaki2018,Stehouwer2018}, which is otherwise hard to compute.

In this work, we focus on the classification and topological invariants for 2D time-reversal superconductors with translational and $C_2$ rotational symmetries. 
The key reason why we choose to focus on this symmetry class is because this is the simplest example where the bulk topology \textit{cannot} be revealed by high-symmetry points and that the topological invariants are \textit{not} symmetry indicators, as we will explicitly show in Sec.~\ref{sec:AHSS} and Appendix A.

The symmetry group of our case study is described as follows. For a given 2D mean-field BdG Hamiltonian at crystal momentum $\kvec$ 
\be
	H(\kvec) = \begin{pmatrix}
	h(\kvec) & \Delta(\kvec) \\
	\Delta(\kvec)^\dagger & -h(-\kvec)^* 
	\end{pmatrix} 
\ee
written in the Nambu basis $\left( \hat{c}_{\kvec,\uparrow}, \hat{c}_{\kvec,\downarrow}, \hat{c}^\dagger_{-\kvec,\uparrow}, \hat{c}^\dagger_{-\kvec,\downarrow} \right)^\trans$ suppressing indices for other degrees of freedom, we consider the case where the single-particle normal state $h(\kvec)$ transforms projectively under the two-fold rotation as $U_{C_2}(\kvec) h(\kvec) U_{C_2}(\kvec)^{-1} = h(-\kvec)$ with $U_{C_2}(-\kvec) U_{C_2}(\kvec) =-1$, 
and the superconducting order parameter $\Delta(\kvec)$ transforms trivially as $U_{C_2}(\kvec) \Delta(\kvec) U_{C_2}(-\kvec)^\trans = \Delta(-\kvec)$.
The symmetry group $G$ of the full BdG Hamiltonian $H(\kvec)$, besides the translational symmetries in $x$ and $y$ directions, contains the time-reversal symmetry $\TR$, particle-hole symmetry $\PH$, and the two-fold BdG rotational symmetry $\CR(\kvec)=\mathrm{diag} \{U_{C_2}(\kvec), U^*_{C_2}(-\kvec)\}$. The BdG Hamiltonian therefore satisfies the symmetry constraints
\begin{subequations} \label{eq:sym_ham}
\begin{align}
	\TR H(\kvec) \TR^{-1} = H(-\kvec), \\
	\PH H(\kvec) \PH^{-1} = -H(-\kvec), \\
	\CR(\kvec) H(\kvec) \CR(\kvec)^{-1} = H(-\kvec),
\end{align}
\end{subequations}
where the symmetries obey the following group relations 
\begin{subequations} \label{eq:sym}
\begin{align}
	\TR^2=-1, \, \PH^2=1, \, \CR(-\kvec)\CR(\kvec)=-1, \\
	\TR \CR(\kvec)=\CR(-\kvec) \TR, \, \PH \CR(\kvec)=\CR(-\kvec) \PH. 
\end{align}
\end{subequations}
We note that the $C_{2}$ rotation $\CR$ squares to $-1$ since we consider spinful electrons. The transformation rule of the pairing function $\Delta(\kvec)$ under $C_2$ rotation determines the form of $\CR$ and thus the commutation relation between $\CR$ and $\PH$. 
Together with the fact that the time-reversal and particle-hole symmetries are anti-unitary such that $\phi(\TR)=\phi(\PH)=-1$, the relations in Eqs.~(\ref{eq:sym_ham}) and (\ref{eq:sym}) fully define the factor system $\phi$, $\tau$, and $c$ of the equivariant K group $^\phi K_G^{(\tau,c),-n}(BZ)$ for 2D superconductors of symmetry group $G$.

\section{Momentum-space classification}
\label{sec:AHSS}

In this section, we will calculate the K group $^\phi K_G^{(\tau,c),-n}(BZ)$ using the AHSS\cite{Shiozaki2018,Huang2021} for 2D time-reversal superconductors with the two-fold rotational symmetry whose rotational axis is out of plane. In the following, we will illustrate the key steps of the formalism and refer the readers to the details in Appendix~\ref{app:AHSS}.  

\subsection{Overview}

We first provide an intuitive overview of this mathematical tool before diving into the formal construction. AHSS approximates the equivariant K group for systems with crystalline symmetries using the following strategy: Instead of computing the K group on the whole BZ directly, which is generally a difficult task in the presence of crystalline symmetries, one first decomposes the BZ into different dimensional subspaces that do not obey any crystalline symmetries which are not local in momentum. 
One can then study the `local K groups' restricted to these subspaces. These local K groups can be interpreted as the classifications of gapped Hamiltonian restricted to different dimensional subspaces, which is dubbed the `topological phenomena interpretation'\cite{Huang2021}. Since these local Hamiltonians only obey symmetries that are local in momentum, the local K groups are effectively given by the classifications of AZ classes.

These independent local K groups together provide the zeroth-order approximation of the whole K group, and is the starting point of the spectral sequence.  
Next, one improves the approximation order by order through proper assembly processes of these local K groups until the approximated result converges, where the assembly processes are guided by the compatibility relations between different dimensional local K groups. 
The AHSS is a formal mathematical tool that encodes all these structures and allows one to systematically consider these compatibility relations.

Practically speaking, it is often convenient for calculation purposes to adopt the `representation interpretation'~\cite{Huang2021} and view the local K groups as the groups associated with the {band irreducible representations (irreps) on high-symmetry points, lines, and general points in the 2D BZ. 
Then, the assembly processes are performed by modifying the representation groups according to, for instance, how many nontrivial representations at a high-symmetry point are not trivialized when extended to the neighboring lines or general points. For our focus of crystalline superconductors, these  representations are from the BdG bands of the considered local BdG Hamiltonians. 

After performing all the assembly processes dimension by dimension, one can obtain the final groups of the representations on the high-symmetry points, lines, and general BZ points.
These representation groups (i.e. local K groups) for different dimensional subspaces can then be characterized by different dimensional topological invariants, such as symmetry indicators, winding numbers, and Chern numbers. 
The purpose of our AHSS calculation presented in this section is to identify which of the local K groups are nontrivial for our case study of 2D superconductors with time-reversal and two-fold rotational symmetries.

\subsection{Cell decomposition}

\begin{figure}[h]
	\centerline{
	\begin{tikzpicture}
		\tikzstyle{zeroc} = [draw, shape=circle, fill=white, minimum size=1.0em]
		\tikzstyle{onec} = [blue,thick,->]
		\node[zeroc] (q00) at (0,0) {$\Gamma$};
		\node[zeroc] (q10) at (2.0,0) {$X$};
		\node[zeroc] (q20) at (-2.0,0) {$X$};
		\node[zeroc] (q01) at (0,2.0) {$Y$};
		\node[zeroc] (q02) at (0,-2.0) {$Y$};
		\node[zeroc] (q11) at (2.0,2.0) {$M$};
		\node[zeroc] (q12) at (2.0,-2.0) {$M$} edge [blue,thick,<-] (q02);
		\node[zeroc] (q21) at (-2.0,2.0) {$M$} edge [blue,thick,<-] (q20);
		\node[zeroc] (q22) at (-2.0,-2.0) {$M$} edge [blue,thick,<-] (q20) edge [blue,thick,<-] (q02);
		\node (c1) at (0,1.0) {$\alpha$};
		\node (c2) at (0,-1.0) {$\alpha^\prime$};
		\draw (q00) [onec] -- node[above] {$a$} ++(q10);
		\draw (q00) [onec] -- node[above] {$a^\prime$} ++(q20);
		\draw (q10) [onec] -- node[right] {$b$} ++(q11);
		\draw (q10) [onec] -- node[right] {$b^\prime$} ++(q12);
		\draw (q01) [onec] -- node[above] {$c$} ++(q11);
		\draw (q01) [onec] -- node[above] {$c^\prime$} ++(q21);
		\draw[red,->] (c1.90) arc (0:240:2mm);
		\draw[red,->] (c2.-90) arc (180:180+240:2mm);
	\end{tikzpicture}
	}
	\caption{The symmetry-preserving cell decomposition for the Brillouin zone in the momentum space. The circles denote the 0-cells, $\{a$, $b$, $c\}$ denote the independent 1-cells, and $\alpha$ denotes the independent 2-cell. $\{a^\prime$, $b^\prime$, $c^\prime\}$ and $\alpha^\prime$ are the $C_2$-related counterparts of $\{a$, $b$, $c\}$ and  $\alpha$, respectively. The arrows on 1-cells and 2-cells represent the orientations of these cells.}
\label{fig:BZ}
\end{figure}
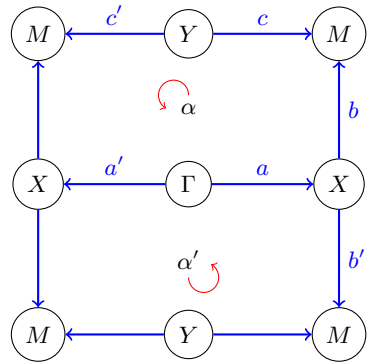

The first step of AHSS is to construct a `cell decomposition' of the 2D BZ in a way such that (1) each symmetry operator in the considered symmetry group $G$ either acts as the identity operation on a given cell or maps it to another cell of the same dimensionality, and (2) the orientations of the cells are consistent among themselves under symmetry actions (see the example in Fig.~\ref{fig:BZ}). A cell of dimensionality $p$ is referred to as a $p$-cell, and all cells related by the symmetries in $G$ together form an orbit. 
For calculation purposes, it is sufficient to consider a representative $p$-cell out of each orbit. 

For our case study, the cell decomposition of the 2D BZ with the symmetry group $G$ is given in Fig.~\ref{fig:BZ}. The 0-cells form four orbits $\{\Gamma\}$, $\{X\}$, $\{Y\}$, and $\{M\}$; the 1-cells form three orbits $\{a, a^\prime\}$, $\{b, b^\prime\}$, and $\{c, c^\prime\}$; the 2-cells form one orbit $\{\alpha, \alpha^\prime\}$. Of each dimension, an independent set of cells that are not related by symmetries consists of one cell from each different orbit. The 0-cells $\Gamma, X, Y, M$ are independent of each other; $a, b, c$ together form an independent set of 1-cells; $\alpha$ is the only independent 2-cell.

\begin{table*}
\begin{center}
\begin{tabular}{c | c | c | c | c | c }
	\hline
	$E_1^{\bar{p},-\bar{n}}$ & ... & $\bar{p}=p$ & $\bar{p}=p+1$ & $\bar{p}=p+2$ & ... \\ \hline
	$\bar{n}=n+p$ & ... & $p$-D gapped Hamiltonians & $(p+1)$-D gapless Hamiltonians & $(p+2)$-D Hamiltonians with singular points & ... \\ \hline
	$\bar{n}=n+p+1$ & ... & ? & $(p+1)$-D gapped Hamiltonians & $(p+2)$-D gapless Hamiltonians & ... \\ \hline
	$\bar{n}=n+p+2$ & ... & ? & ? & $(p+2)$-D gapped Hamiltonians & ... \\ \hline
	$\vdots$ & & $\vdots$ & $\vdots$ & $\vdots$ & \\ \hline
\end{tabular}
\end{center}
\caption{The topological phenomena interpretation of the $E_1$ page entries in the AHSS for classifying systems in the AZ class $n$. We leave question marks in the entries whose physical meanings are not clear in this interpretation. Nonetheless, all entries are well-defined and can be calculated in the representation interpretation (see the main text).}
\label{tab:AHSS}
\end{table*}

\subsection{Pages and differentials}

Given the cell decomposition, the second step is to study the local K groups restricted on different dimensional cells and the compatibility relations among them. 
In the language of AHSS, the first page $E_1$ is a table that consists of the local K groups (without considering any compatibility relations) bigraded by the dimensionality $p$ and the AZ symmetry class $n$ \cite{Shiozaki2018}.
The physical meaning of the table can be understood by the topological phenomena interpretation:  
Each entry $E_1^{p, -(n+p)}$ is an Abelian group that characterizes the classification of \textit{gapped} systems with the internal symmetries given by the AZ class $n$ and the $k$ space as disjoint $p$-spheres\cite{E1}. Each $p$-sphere is a representative $p$-cell with its boundary identified to a point. $E_1^{p+1, -(n+p)}$ can be understood as the classification of \textit{gapless} systems on the $(p+1)$-cells with the internal symmetries given by the AZ class $n$. For clarity, we summarize the physical meaning of the $E_1$ page entries in Table~\ref{tab:AHSS} under the topological phenomena interpretation. The $E_1$ page for our case study is given in Table~\ref{tab:E1}.

Next, we assemble the local K groups on $p$-cells with those on $p\pm1$-cells by incorporating the relevant compatibility relations. This is formally done by studying the first differential $d_1^{p, -(n+p)}:E_1^{p, -(n+p)} \rightarrow E_1^{p+1, -(n+p)}$, which is a homomorphism from every entry in the $E_1$ page to its right neighboring entry on the same page. 
In the topological phenomena interpretation, this map represents symmetry-allowed processes that extend a gapped Hamiltonian on a $p$-cell to a gapless Hamiltonian on the adjacent $(p+1)$-cells in the same AZ class $n$. Examples of such processes include a shift in the chemical potential or a band inversion~\cite{Shiozaki2018}. 

By taking into account the assembly processes described by the first differentials, the local K groups are now modified into  $E^{p,-(n+p)}_{2} \equiv {\rm Ker}(d^{p,-(n+p)}_1)/{\rm Im}(d^{p-1,-(n+p)}_1)$ and are organized into the table of $E_2$ page. 
It is most intuitive to understand this expression for some entries in the $E_2$ page in the \textit{topological phenomena} interpretation. Let us consider the entry $E_2^{0,-3}$ (see Table~\ref{tab:E2}) as an example. Recall that the diagonal entry $E_1^{0,-3}$ and the off-diagonal entry $E_1^{1,-3}$ in the $E_1$-page are 
the classifications of class-DIII gapped Hamiltonians on 0-cells and gapless Hamiltonians on 1-cells in the same class, respectively. Therefore, to improve the approximation of the diagonal entries $E_1^{0,-3}$ by considering the compatibility relation described by the first differential $d_1^{0,-3}$, we keep only the gapped Hamiltonians on 0-cells in $E_1^{0,-3}$ that remain gapped on 1-cells. This set of Hamiltonians is given by Ker($d_1^{0,-3}$). We therefore arrive at $E^{0,-3}_{2}={\rm Ker}(d^{0,-3}_1)$, where the differential $d^{-1,-3}_1$ is trivial.

Nonetheless, to fully explain the expression for general entries $E^{p,-(n+p)}_{2}$ in the $E_2$-page, it is most transparent and calculation-wise convenient to turn to the \textit{representation} interpretation since the lower half of the $E_1$-page does not have clear physical meaning in the topological phenomena interpretation (labeled by the question marks in Table \ref{tab:AHSS}). The two interpretations are mathematically equivalent \cite{Shiozaki2018}. 
In the representation interpretation, the $E_1$ page entry $E^{p,-(n+p)}_1$ classifies the irreps of the effective symmetry group for a point within each $p$-cell with internal symmetries in the AZ class $n+p$. The first differential $d^{p,-(n+p)}_1$ describes the process of mapping an irrep on a $p$-cell to one or several irreps of the smaller effective symmetry group for a point in an adjacent $(p+1)$-cell. In the next step of approximating contribution from the $p$-cells, only these irreps in ${\rm Ker}(d_1^{p,-(n+p)})$ remain because others correspond to independent irreps on the adjacent $(p+1)$-cells. On the other hand, the irreps in ${\rm Im}(d_1^{p-1,-(n+p)})$ can be trivialized by the irreps on the adjacent $(p-1)$-cells so they do not contribute. Trivialization here means the process of moving irreps on adjacent lower dimensional-cells to a higher-dimensional cell and forming a trivial irrep together with the original irrep on that cell. For detail about the interpretations, we refer the readers to pedagogical discussions in Refs. \cite{Shiozaki2018,Huang2021}.  

Following similar procedures of obtaining the $E_2$-page by considering the first differentials, we can take into account higher-order assembly processes, which are described generally by the $r$-th differential $d_r^{p,-(n+p)}:E_r^{p, -(n+p)} \rightarrow E_r^{p+r, -(n+p+r-1)}$, order by order to obtain better approximated local K groups 
\begin{align}
	E_{r+1}^{p, -(n+p)} = {\rm Ker}(d_r^{p,-(n+p)}) / {\rm Im}(d_r^{p-r,-(n+p-r+1)}). 
	\label{eq:Er}
\end{align}
A higher page $E_{r+1}$ therefore provides a better approximation to the K group than the previous page $E_r$. By following Eq.~\ref{eq:Er} to iteratively improve the approximation of the local K groups, the $E_r$ page will eventually converge at some $r^*$ (i.e. $E_{r^*}=E_{r^*+r^\prime}\,\forall r^\prime\geq1$). Such convergence will happen when $r\geq d+1$ or before that for a $d$-dimensional system since $d_r=0$ for any $r\geq d+1$. 
This final converged page is dubbed the limiting page $E_{\infty}$. 

For our case study, the $E_1$ page and the infinity page $E_\infty$ are given in Table~\ref{tab:E1} and Table~\ref{tab:E2}. The detailed calculation of how to arrive at these tables is also given in Appendix~\ref{app:AHSS}.

\begin{table}[h]
\begin{center}
\begin{tabular}{c c | c | c | c}
	\hline
	AZ class & $\bar{n}$ & $\bar{p}=0$ & $\bar{p}=1$ & $\bar{p}=2$ \\ \hline
	DIII & 3 & $1$ & $1$ & $1$ \\ \hline
	AII & 4 & $\zee^4$ & $\zee^3$ & $\zee$ \\ \hline
	CII & 5 & $1$ & $\ztwo^3$ & $\ztwo$ \\ \hline
\end{tabular}
\end{center}
\caption{The $E_1$ page we find from our calculation.}
\label{tab:E1}
\end{table}

\begin{table}[h]
\begin{center}
\begin{tabular}{c c | c | c | c}
	\hline
	AZ class & $\bar{n}$ & $\bar{p}=0$ & $\bar{p}=1$ & $\bar{p}=2$ \\ \hline
	DIII & 3 & $1$ & $1$ & $1$ \\ \hline
	AII & 4 & $\zee$ & $\ztwo^3$ & $\zee$ \\ \hline
	CII & 5 & $1$ & $\ztwo^3$ & $\ztwo$ \\ \hline
\end{tabular}
\end{center}
\caption{The $E_2$ page we find from our calculation. The $E_2$ page is also the limiting page $E_\infty$ since it is equal to all higher pages. According to the topological phenomena interpretation, the three diagonal entries represent the K groups restricted on 0-, 1-, and 2-cells for the 2D class-DIII superconductors with $C_2$ symmetry. These three entries together give rise to the full K group.}
\label{tab:E2}
\end{table}

\subsection{Topological invariants}
\label{sec:inv}

We now discuss how to extract information of topological invariants for the full system from the infinity page. 
In the topological phenomena interpretation, each of the diagonal entries in the limiting page $E_{\infty}^{p, -(n+p)}, \, 0 \leq p \leq d$ represents the classification of gapped systems on the $p$-dimensional subspaces of the BZ for the AZ class $n$.  
Moreover, these diagonal entries are subgroups of the full K group and can lead to the full K group through a short exact sequence with certain group extensions (see Appendix~\ref{app:AHSS}). 
These diagonal entries can therefore be characterized by a set of topological invariants on different dimensional subspaces that encodes the topology of the full system. 

There are three pieces of information about the invariants that we can extract from the diagonal entries $E_{\infty}^{p, -(n+p)}$. First, these groups directly tell us the type of the topological invariants that are necessary to capture the topology of the full system. Second, these invariants take band data as inputs and we will know from the diagonal entries which dimensional subspace in the BZ contributes to each invariant. Finally, the information of each invariant is in the form of a homotopy group of some topological space, which prompts us to find an invariant characterizing the maps from a subspace of the BZ to that topological space.

For our case study of 2D class-DIII superconductors with a two-fold rotational symmetry, there are three diagonal entries in the infinity page:   $E_{\infty}^{0,-3}$, $E_{\infty}^{1,-4}$, and $E_{\infty}^{2,-5}$. These three entries are the local K groups restricted on the 0-cells $\{\Gamma$, $X$, $Y$, $M$\}, the 1-cells $\{a,b,c\}$, and the 2-cell $\alpha$, respectively. 
We find that the three subgroups are (see Table~\ref{tab:E2})
\begin{align}
	E_{\infty}^{0,-3}=1, E_{\infty}^{1,-4}=(\ztwo)^3,  E_{\infty}^{2,-5}=\ztwo,  
\end{align}
where the $(\ztwo)^3$ comes from one $\ztwo$ per 1-cell. 
We therefore expect that the topology of the full system cannot be inferred from the band information on the high symmetry points (as $E_{\infty}^{0,-3}=1$) but can be fully characterized by four $\ztwo$ invariants, three depending on the band data on the \textit{high-symmetry lines} and one on \textit{general points} in the BZ.

Importantly, we can further deduce the topological spaces for the two types of $\ztwo$ invariants as follows. First, as we show in Appendix~\ref{app:AHSS}, the two nontrivial local K groups $E_{\infty}^{1,-4}$ and $E_{\infty}^{2,-5}$ are essentially obtained from $\pi_0(R_0)=\zee$ and $\pi_0(R_1)=\ztwo$, respectively, where  $R_0=O(N_1+N_2)/(O(N_1)\times O(N_2))$ and $R_1=O(N)$. Here $N$, $N_1$, $N_2$ are integers that should be taken to the infinite limit. To see why $E_{\infty}^{1,-4}=(\ztwo)^3$, we note that there is a modulo 2 operation produced by identifying ${\rm Im}(d_1^{0,-4})$ as trivial when calculating the $E_2$ page out of the $E_1$ page. We will see this in terms of the definition for the invariant in Sec.~\ref{sec:1d}. 
Moreover, the Bott periodicity theorem~\cite{Wen2012,Ludwig2015} leads to the following identities
\be
	\pi_0(R_0)=\pi_1(R_7),  \, \pi_0(R_1)=\pi_2(R_7),
	\label{eq:Bott}
\ee
where $R_7=U(N)/O(N)$. We therefore expect that the topological spaces for the two types of $\ztwo$ invariants should both be the space $R_7$. 

In summary, our AHSS results suggest that the topology of a 2D class-DIII Hamiltonian with a two-fold rotational symmetry cannot be inferred from band data on the high symmetry points, but can be fully characterized by three $\ztwo$ invariants defined on the high symmetry lines and one $\ztwo$ invariant on general points in the BZ. 
In particular, we expect that valid forms of the two types of $\ztwo$ invariants can be constructed by maps from 1-cells to the space $R_7$ and 2-cells to $R_7$, respectively. 
In the rest of the paper, we will propose specific forms for these topological invariants and show that the set of invariants can diagnose the Majorana boundary type by establishing the bulk-boundary correspondence.

\section{Explicit expressions for topological invariants}
\label{sec:sew} 

From our AHSS calculation, we find that only the K groups restricted on 1-cells and 2-cells are nontrivial, and our goal is to write down topological invariants that can fully distinguish the phases. 
In fact, they have readily provided us hints for the forms of these invariants. Specifically, the K groups restricted on 1-cells and 2-cells are given by $\pi_1(R_7)/{\rm Im}(d_1^{0,-4})=\ztwo$ and $\pi_2(R_7)=\ztwo$ [see Eq.~(\ref{eq:Bott})], which are based on the first and second homotopy groups of space $R_7=U(N)/O(N)$, respectively.
Therefore, to write down valid 1D and 2D invariants that characterize the full bulk topology, our strategy is to find an object that lives in the space $R_7$ and takes the symmetry band data on 1- and 2-cells as inputs, respectively. The 1D and 2D invariants are then given by the first and second homotopy classes of this object. 

We propose to use the sewing matrix 
\be
	\sew_{mn} (\kvec) \equiv \bra{u_m(\kvec)} \TR \CR \ket{u_n(\kvec)}
	\label{eq:sew} 
\ee
of the combined symmetry $\TR \CR$ to construct the object, where $\TR$ is the time reversal symmetry, $\CR$ is the two-fold rotational symmetry, and $\ket{u_n(\kvec)}$ is the $n$-th occupied state at momentum $\kvec$.

To see why $\sew$ lives in the space of $R_7$, where $R_7=U(N)/O(N)$ for $N\rightarrow\infty$, first notice that $\sew$ is a symmetric matrix since $\sew$ is an anti-unitary operator that is local in $\kvec$ and satisfies $(\TR \CR)^2=1$.
As a symmetric matrix, $\sew$ has a (non-unique) decomposition $\sew(\kvec)=U(\kvec)U(\kvec)^\trans$ where $U(\kvec)$ is unitary, known as Autonne-Takagi factorization~\cite{Takagi1924,Horn2012,Dai2020}. $\sew$ therefore remains invariant under the change $U \rightarrow UO$ for any orthogonal matrix $O$. The number of occupied bands can be taken to the infinite integer by adding any number of trivial bands since the K theory classification is stable against this operation. Therefore, $\sew$ lies in $R_7$. 
Moreover, since there are even number $2N_0$ of occupied states due to $\TR^2=-1$, we can always consider half of the time reversal related pairs, i.e. $N_0$ states that are not mapped to each other by time reversal. We label this set of occupied states and its time reversal related partner by $I$ and $II$ respectively. This corresponds to block diagonalizing $\sew$ into $\sew_{I}$ and $\sew_{II}$, where the two blocks are related by $\TR$, and then taking only one of them. $\sew_{I}$ (or equivalently $\sew_{II}$) thus defines a map from the BZ to the space of $U(N_0)/O(N_0)$. In the limit of a large number of occupied bands, the target space of $\sew_{I/II}$ is also $R_7$.

Importantly, since the sewing matrix depends on the occupied wavefunctions, $\sew$ is gauge-dependent. Specifically, under a general gauge transformation $V$, $\sew$ transforms as
\be
	\sew \rightarrow V^\dagger \sew V^*.
	\label{eq:gauge}
\ee
There are two gauge choices that are relevant for the invariants we propose: the first is dubbed the $\TR$-constant and smooth gauge and the second is dubbed the real gauge.

We first define the $\TR$-constant and smooth gauge. In a subspace of the BZ, this gauge satisfies both of the following 
\begin{enumerate}
	\item Smooth: the wavefunctions are smooth;
	\item $\TR$ constant: the matrix $\bra{u_m(-\kvec)} \TR \ket{u_n(\kvec)}$ is independent of $\kvec$.
\end{enumerate}
We refer to the first and the second conditions as `smooth' and `$\TR$-constant', respectively. 
The $\TR$-constant condition was also used in one definition for the Fu-Kane invariant~\cite{Fu2006,Teo2010} and the closely related first $\ztwo$ descendent for the chiral classes in Ref.~\cite{Chiu2016}. Loosely speaking, it allows us to separate the time-reversal-related partners and focus on one block $\sew_{I/II}$. There may be obstruction to satisfying both of the gauge conditions everywhere in the BZ, which suggests we can no longer consider only one of the time-reversal-related partners.

In fact, these conditions alone do not fix a unique gauge---there exist other $\TR$-constant and smooth gauges, which are related to each other by a smooth gauge transformation. All such gauges are considered equivalent as far as the $\sew$-based invariants are concerned. In particular, as we will discuss in detail in Sec.~\ref{sec:1d} and Appendix~\ref{app:AHSS}, this gauge freedom corresponds to the normal subgroup ${\rm Im}(d_1^{0,-4})$ of $\pi_{1}(R_{7})$ so that the classifying group of the 1D invariants is given by the quotient group $\pi_1(R_7)/{\rm Im}(d_1^{0,-4})=\ztwo$. In other words, for one of the 1D invariants expressed as an integer in the group $\pi_1(R_7)=\zee$, calculating this invariant in another smooth and $\TR$-constant gauge modifies it by an even integer.

Next, we define the real gauge. If we carry out a gauge transformation $V=U(\kvec)$, the sewing matrix will become an  identity $\sew=\iden$ under this gauge (see Eq.~(\ref{eq:gauge})). We refer to this gauge as the real gauge~\cite{Ahn2018,Ahn2019,RealGauge}. 
As we will discuss in Sec.~\ref{sec:1d} to ~\ref{sec:SW}, when we calculate our proposed invariants by taking the homotopy classes of $\sew_I$ (or equivalently $\sew_{II}$), we choose the \textit{smooth and $\TR$-constant gauge} for the sewing matrices. On the other hand, when we interpret the invariants as the S-W classes, we calculate the occupied wavefunctions in the \textit{real gauge}.

We now make a final remark about this sewing matrix before moving on. In fact, sewing matrices of time reversal and the combined operation of time reversal and $C_2$ rotation have been used to construct the topological invariants in several insulating systems~\cite{Fu2006,Ahn2018,Ahn2019}.
Here in our case study of superconductors, we choose to use the sewing matrix of $\TR \CR$ instead of other symmetry combinations, such as $\PH \CR$ or $\PH \TR$, because the spaces of those sewing matrices are not $R_7$. Specifically, $\PH \CR \in U(2N)/Sp(N)$ and $\PH \TR \in U(2N)/U(p,q)$ where $p+q=2N$.

\subsection{Reference Hamiltonian}
\label{sec:refH}

Before we write down the actual expressions for the 1D and 2D invariants based on the homotopy groups of the sewing matrix $\sew$, we first revisit some properties of elements in a K group and the forms of valid invariants that are capable of distinguishing these elements. In particular, we will see that it is necessary to introduce a universal reference Hamiltonian to the analysis and incorporate it in the invariants accordingly.  

In K theory~\cite{Shiozaki2017,Shiozaki2018,Stehouwer2018,Huang2021}, a K group can be viewed as consisting of adiabatic paths between two gapped Bloch Hamiltonians $H_1$ and $H_2$ living in some Hilbert space, which obeys symmetries in a given symmetry class. These paths can be grouped into different equivalent classes following equivalence relations expected from physics grounds~\cite{Shiozaki2017,Shiozaki2018,Stehouwer2018,Huang2021}, and the K group classifies how many inequivalent classes of path there are for the given symmetry class. 
In Karoubi's formulation, such a class of paths is mathematically denoted as a triple $[\VB, H_1, H_2]$~\cite{Karoubi1978,Shiozaki2017,Shiozaki2018,Stehouwer2018,Huang2021}, where $\VB$ is the vector bundle whose base space is the BZ and vector space is formed by the occupied states of the Hamiltonian $H_1$. Here, the path connecting $H_1$ and the reference Hamiltonian $H_2$ in $\VB$ is a representative path of the equivalence class.   

We can further associate different equivalence classes of paths $[\VB, H, H_0]$ with topologically distinct classes of gapped phases of matter. 
To do so, instead of using different reference Hamiltonians $H_2$ for different triples, it is crucial to define a `trivial' Hamiltonian $H_0$ as the universal reference Hamiltonian (i.e. to set $H_2=H_0$) for all triples. 
For our purpose of investigating superconductors, we define the trivial Hamiltonian to be the BdG Hamiltonian 
\begin{align}
  H_0 = \rm{diag}(\mathbb{I}_N, -\mathbb{I}_N),  
  \label{eq:H0}
\end{align}
formed by a normal vacuum state, which is described by a normal-state Hamiltonian consisting of $N$ flat bands with a chemical potential set at negative infinity. 
Under such a construction, each Hamiltonian $H$ belongs to the phase represented by the triple $[\VB, H, H_0]$, and the triple $[\VB, H_0, H_0]$ is the identity element in the K group. It is worth noting that under different symmetry actions implicitly specified by $\VB$ and $\VB^\prime$, the triples $[\VB, H, H_0]$ and $[\VB^\prime, H, H_0]$ can represent different phases even though the Hamiltonians are in the same form $H$ and $H_0$. 

Given the K group of the considered symmetry class, now the goal is to construct the topological invariants that are capable of distinguishing Hamiltonians that belong to topologically inequivalent phases. 
Specifically, we are after the phase label $\nu(\VB,H,H_0)$, which should take different values for triples belonging to different equivalent classes. Nonetheless, it is practically hard to write down the actual form for $\nu$. We therefore turn to writing down invariants $w(\epsilon,h)$ defined for a given Hamiltonian $h$ living in the vector bundle $\epsilon$ (instead of a given triple). 
The key difficulty in writing down $w$ occurs when the symmetry operators carry momentum dependence, which is implicitly encoded in the vector bundle $\epsilon$. In such cases, the topological invariant $w$ is generally non-zero for the trivial Hamiltonian $h=H_0$ defined in Eq.~(\ref{eq:H0}). 
This could therefore lead to confusion when identifying topological phases because $H_0$ necessarily corresponds to a trivial phase even when $w(\epsilon,H_0)\neq 0$ due to the nontrivial vector bundle $\epsilon$ (shown by the real-space classification in Sec.~\ref{sec:real_class}).

We propose to resolve this issue in the following way. 
To account for the contribution to the invariant $w$ from the reference Hamiltonian $H_0$, we need to find a Hamiltonian $H_c$ acting on a vector bundle $\VB_c$ such that all the invariants vanish for $H_0 \oplus H_c$, i.e. $w(\VB \oplus \VB_c,H_0\oplus H_c)=0$. Such $H_c$ is $H_0$-dependent, and should cancel out the implicit $H_0$ contribution to $w$ even when the symmetry operators are momentum-dependent. 
This implies that when calculating the topological invariant $w$ for a given Hamiltonian $H$ of interest, the correct quantity to calculate is $w(\VB \oplus \VB_c, H \oplus H_c)$, which guarantees that $w=0$ when $H$ is trivial for any choice of $\VB$. In other words, we have identified this practically calculable quantity as the phase label defined for the corresponding triple 
\begin{align}
	\nu(\VB,H,H_0):=w(\VB \oplus \VB_c, H \oplus H_c).
	\label{eq:nuw}
\end{align}
This quantity $w(\VB \oplus \VB_c, H \oplus H_c)$ is what we will write down and compute for the 1D and 2D invariants for our case study in the following subsections.  

To provide some examples for the choice of $H_c$, when the invariants $w$ of a given system are symmetry indicators, it has been proposed that the reference Hamiltonian $H_0$ should be accounted for by subtracting the symmetry indicators of $H_0$~\cite{Shiozaki2019,Skurativska2020,Ono2020,Huang2021}. This is equivalent to setting $H_c=-H_0$ in our approach. 
Another example is our current case of 2D class-DIII superconductors with $C_2$ rotational symmetry, where the invariants $w$ are \textit{not} symmetry indicators but 1D and 2D invariants. We will show in Sec.~\ref{sec:1d} and ~\ref{sec:2d} that we can generally choose $H_c=H_0$ for the $\ztwo$ invariants we write down.

\subsection{One-dimensional invariants}
\label{sec:1d}

In this subsection, we propose an explicit form for the $\ztwo$ invariants $w_1$ defined on 1-cells, which can distinguish phases in the subgroup $E_{\infty}^{1,-4}$ of the K group.  
We write down this form of $w_1$ based on our AHSS result, which shows that the contribution to the total K group from the three independent 1-cells $\{a,b,c\}$ is three copies of $\ztwo$, each obtained from $\pi_1(R_7)/\text{Im}(d_1^{0,-4})=\ztwo$. 
The key idea here is to consider the sewing matrix $\sew$ [see Eq.~(\ref{eq:sew})] and the derived $\sew_{I/II}$, both restricted to the 1-cells, since they live in the space of $R_7$ as the number of bands goes to infinity.    
The first homotopy class of $\sew_{I/II}$ in general takes all integer values. Nonetheless, when requiring this quantity to be gauge-invariant under the $\TR$-constant and smooth gauge conditions, under which $\sew_{I/II}$ is well-defined, the redundant gauge degrees of freedom included in Im($d_1^{0,-4}$) can be removed and the first homotopy class of $\sew_{I/II}$ becomes a $\ztwo$ quantity.
We therefore propose to write down each of the $\ztwo$ invariant $w_1$ as the first homotopy class of $\sew_{I/II}$ under the $\TR$-constant and smooth gauge. 
These three $\ztwo$ invariants defined on the three 1-cells together can diagnose whether a given superconductor is in the first-order strong topological superconducting phase, and we will discuss how to predict the actual boundary signatures in Sec.~\ref{sec:real}.

Specifically, for a given BdG Hamiltonian, we start from the quantity 
\be
	w_{1j} = \frac{1}{2\pi} \int_j dk \, \pt_k \log\det \sew \, \mod 2, \, j=a,b,c,  
	\label{eq:w1}
\ee
which is the line integral of the phase of  
the sewing matrix $\sew$ [see Eq.~(\ref{eq:sew})] along 1-cell $j$. 
Here, $\sew$ is calculated under the $\TR$-constant and smooth gauge conditions, where $\sew$ is well-defined and can be block-diagonalized into $\sew=\text{diag}(\sew_{I},\sew_{II})$ with $\sew_{I}$ and $\sew_{II}$ being the time-reversal partner of each other. 

Under this gauge, we can further express $w_{1j}$ as a loop integral over only one of the time-reversal copies in the sewing matrix  
\begin{align}
	w_{1j}&=\frac{1}{2\pi} \oint_{l(j\cup j^\prime)} dk \, \pt_k \log\det \sew_{I}\nonumber \\ 
	&= \frac{1}{2\pi} \oint_{l(j\cup j^\prime)} dk \, \pt_k \log\det \sew_{II}, 
	\label{eq:wind}
\end{align}
where $l(j\cup j^\prime)$ denotes the loop formed by a 1-cell $j \in \{a,b,c\}$, its symmetry-related counterpart $j^\prime$, and the boundary between them (see Fig.~\ref{fig:BZ}), and the second equality is true since the $\TR$-constant condition implies $\pt_k \log\det \sew_{I}(k) = -\pt_k \log\det \sew_{II}(-k)$. 
Importantly, it is clear from Eq.~(\ref{eq:wind}) that $w_{1j}$ is the winding number of the phase of $\sew_{I/II}$ over the 1-cell loop, which characterizes the first homotopy class of $\sew_{I/II}$. 

Since $\sew_{I/II}$ lives in the space $R_7$, $w_{1j}$ corresponds to the group elements in $\pi_1(R_7)=\zee$~\cite{Hall2015,Ahn2019} and takes all integer values. 
Nonetheless, by requiring the winding number $w_{1j}$ to be gauge invariant under the $\TR$-constant and smooth gauge condition, $w_{1j}$ becomes a $\ztwo$ invariant after the redundant gauge degrees of freedom are removed. 
To see this, first note that when $\sew_{I/II}$ undergoes a gauge transformation in Eq.~(\ref{eq:gauge}) with a general transformation matrix $U(\kvec)$, the integrand $\log\det \sew_{I/II}$ in Eq.~(\ref{eq:wind}) becomes $\log\det \sew_{I/II} - 2\log\det U$. 
When we impose the $\TR$-constant and smooth gauge conditions both before and after the transformation, according to Eq.~(\ref{eq:wind}), $w_{1j}$ changes by a multiple of $2$. We therefore find that $w_{1j}$ should be a $\ztwo$ quantity, and this redundant degree of freedom within the $\TR$-constant and smooth gauge conditions necessitates the modulo $2$ operation in Eq.~(\ref{eq:w1}).\\ 

After establishing that the quantity $w_{1j}$ is $\ztwo$, we now proceed to the last step for writing down the 1D invariant for our case study. This last step is to remove the unwanted hidden contribution to $w_{1j}$ for a given BdG Hamiltonian $H$ from the universal reference Hamiltonian $H_0$ defined in Eq.~(\ref{eq:H0}). 
Such a contribution is generally non-zero when the combined symmetry $\TR\CR$ is momentum-dependent. 
This issue can be resolved by revisiting the relation in Eq.~(\ref{eq:nuw}) between the fundamental phase label $\nu$ defined for a given triple $[\VB, H, H_0]$ and the calculable invariant $w$ defined for a given Hamiltonian $H$ living in the vector bundle $\VB$. 
Applying our general discussion in Sec.~\ref{sec:refH} to the 1D invariant $w_{1j}$, we can further write Eq.~(\ref{eq:nuw}) into 
\begin{align}
	\nu_{1j}(\VB, H, H_0) &= w_{1j}(\VB \oplus \VB_c, H \oplus H_c) \nonumber \\
		&= w_{1j}(\VB,H)+w_{1j}(\VB_c,H_c),
		\label{eq:nuw2}
\end{align}
where $\VB_c$ and $H_c$ are chosen such that $w_{1j}(\VB \oplus \VB_c, H_0 \oplus H_c)=0$. 
The second line follows from the fact that the sewing matrix for $H \oplus H_c$ can be written as the direct sum of those for $H$ and $H_c$, which is clear from Eq.~(\ref{eq:sew}).  
For the case of $w_{1j}$, we choose $H_c=H_0$ because $w_{1j}$ is a $\ztwo$ quantity. 
Therefore, we conclude that the 1D invariant that we should calculate for a given system is 
\begin{align}
	\nu_{1j}(\VB, H, H_0) = w_{1j}(\VB,H)+w_{1j}(\VB,H_0),  
		\label{eq:nuw3}
\end{align}
which guarantees to remove the contribution from the reference Hamiltonian $H_0$ implicitly hidden in $w_{1j}(\VB,H)$ regardless of the momentum-dependence of the symmetry operator $\TR\CR$. 

Finally, we make a remark about the relation between our $\ztwo$ invariant $w_{1j}$ and the Berry phase. 
Berry phase $\gamma$ is a well-known quantity that captures the topological properties in various systems. It is defined as the path integral $\gamma=\int_{C} d\textbf{k}\cdot A_{\textbf{k}}$ (mod $2\pi$) of the Berry connection $A_{\textbf{k}}=\langle u_{\textbf{k}}|i\nabla_{\textbf{k}}|u_{\textbf{k}}\rangle$ for an eigenstate $|u_{\textbf{k}}\rangle$ at momentum $\textbf{k}$, and is quantized when the path $C$ is a closed loop. 
When the combined symmetry $\TR\CR$ of time-reversal and two-fold rotation is momentum-independent, similar to what was shown in Ref.~\cite{Ahn2018}, the invariant $w_{1j}$ [see Eq.~(\ref{eq:wind})] can be related to the Berry phase as $w_{1j}=\gamma/\pi$. 
When the combined symmetry $\TR\CR$ is momentum-dependent, by removing the possibly non-zero contribution from the universal reference Hamiltonian $H_0$ following Eq.~(\ref{eq:nuw2}), we can arrive at a similar relation $w_{1j}(\VB,H)+w_{1j}(\VB,H_0)=\gamma/\pi$.

\subsection{Two-dimensional invariant}
\label{sec:2d}

In this subsection, we propose a way to compute the $\ztwo$ invariants $\nu_2$ defined on 2-cells, which can distinguish phases in the subgroup $E_{\infty}^{2,-5}$ of the total K group. This expression of $w_2$ that we find is based on our AHSS result, which shows that the contribution to the total K group from 2-cells is $\pi_2(R_7)=\ztwo$. Similar to the $w_1$ case, it is tempting to identify the second homotopy class of the sewing matrix as the invariant $w_2$ since the sewing matrix lives in the space $R_7$. 
However, in contrast to the case of $w_1$, which is clearly the winding number of the sewing matrix along a 1-cell loop, it is not clear how to explicitly calculate the second homotopy class of the sewing matrix defined on 2-cells. 
In the following, we will show how to construct an expression for $w_2$ by mapping this quantity to a transition function defined on a 1D path, which is calculable under certain cases that have no obstruction to the $\TR$-constant and smooth gauge condition over the 2-cells.
This $\ztwo$ invariant $w_2$, as we will see in Sec.~\ref{sec:real}, can
diagnose the higher-order topological superconducting phase protected by the two-fold rotational symmetry.

For the purpose of mapping the second homotopy class of the sewing matrix to a more tangible quantity, we invoke two isomorphisms\cite{Ahn2019,AT}
\begin{align}
    \pi_2(U(N)/O(N)) \cong \pi_2(U(N),O(N)) \cong \pi_1(O(N)),
    \label{eq:iso}
\end{align}
where $\pi_1(O(N))$ is the first homotopy group of $O(N)$, and  $\pi_2(U(N),O(N))$ is the second relative homotopy group for which the interior and the boundary of the 2D base space map to the target spaces of $U(N)$ and $O(N)$, respectively. 
We will now apply Eq.~(\ref{eq:iso}) to a given BdG Hamiltonian with $2N_0$ occupied states from our case study. 

For the left-most term in Eq.~(\ref{eq:iso}), we restrict ourselves to the cases where the $\TR$-constant and smooth gauge conditions are imposed over the 2-cells. 
Under such gauge conditions, the sewing matrix $\sew$ is well-defined everywhere on the BZ and can be block-diagonalized into two time-reversal copies $\sew_{I}$ and $\sew_{II}$.  
We can therefore interpret an element in the left-most term as the second homotopy class of the sewing matrix $\sew_{I/II}$ defined on 2-cells since $\sew_{I/II}$ lives in the space $U(N)/O(N)$ under such gauge conditions. This thus corresponds to our target invariant $w_2$. 

Next, by forming a concrete construction through the middle term in Eq.~(\ref{eq:iso}), we show that the right-most term naturally corresponds to a well-defined quantity that we can calculate for $w_2$. 
This can be most conveniently done under a construction proposed in Ref. \cite{Ahn2019} for insulators with the combined symmetry of time reversal and $C_2$-rotation, which we describe as follows. The first step is to construct the base space of the homotopy group $\pi_2(U(N)/O(N))$ by `distorting' the space formed by the 2-cells and their boundary (i.e. the BZ) to a 2-sphere. 
To this end, first notice that since the 1D invariant $w_1$ [see Eq.~(\ref{eq:wind})] is a $\ztwo$ quantity, we can always find a non-contractible loop $l_1$ (or the sum of the two non-contractible loops) on the 2-torus along which $w_1$ is trivial. 
Since $w_1=0$, the sewing matrix $\sew_{I/II}$ defined along this loop $l_1$ is equivalent to a $\kvec$-independent sewing matrix  $\sew_{I/II}$ (whose $w_1$ is clearly also $0$) in the sense that they are related by a gauge transformation [see Eq.~(\ref{eq:gauge})] within the $\TR$-constant and smooth gauge conditions. 
We can therefore cut the BZ open along $l_1$ and safely identify all the momentum points on each of the two open edges, which all map to the same constant sewing matrix $\sew_{I/II}$, as a point. 
This operation produces a 2-sphere, as illustrated in Fig.~\ref{fig:w2}. 

The second step is to transform the sewing matrix $\sew_{I/II}(\kvec)$ under the smooth gauge, which can be factorized into $\sew_{I/II}(\kvec)=U(\kvec)U(\kvec)^T$ using some unitary matrix $U(\kvec)$, into the real gauge such that $\sew_{I/II}(\kvec)=\mathbb{I}$ for all $\kvec$ on the 2-sphere. Specifically, this can be done by Eq.~(\ref{eq:gauge}) with the transformation matrix $V(\kvec)=U(\kvec)$. 
Under this transformation, although $\sew_{I/II}(\kvec)$ simplifies into the identity, the occupied wavefunctions may have to sacrifice the smooth condition and suffer discontinuity at some $\kvec$'s. 

For simplicity, consider the case where the occupied wavefunctions in the real gauge are everywhere smooth but suffer discontinuities across some closed loop $l_2$. 
In this case, the real-gauge wavefunctions $\ket{u_{L}(\kvec)}$ and $\ket{u_{R}(\kvec)}$ defined on the two hemispheres $\kvec\in L$ and $R$ [see Fig.~\ref{fig:sphere}] are related by a transition function matrix  $O_1(\kvec)$ defined on the loop $l_2$ 
\be
	O_{1, mn}(\kvec) = \braket{u^\prime_{L,m}(\kvec)}{u^\prime_{R,n}(\kvec)}, \, \kvec \in l_2. 
\ee
Here, $m,n=1,\cdots,N_0$ are band labels for the occupied wavefunctions, and the prime symbol ${'}$ in $|u^\prime\rangle$ indicates that the wavefunction is in the real gauge. 
Importantly, since the wavefunctions are in the real gauge, the transition function matrices $O_1(\kvec)$ along the closed loop $\kvec\in l_2$ are orthogonal and belong to the space $O(N)$. 

This discontinuity carried by the real-gauge wavefunctions is also embedded in the transformation matrix $U(\kvec)$, which transforms the sewing matrix $\sew_{I/II}(\kvec)$ from the smooth to the real gauge. 
Specifically, suppose the wavefunctions transform from the smooth to real gauge on each of the two hemispheres $L/R$ as $|u_{L/R}(\kvec)\rangle\rightarrow |u^\prime_{L/R}(\kvec)\rangle=U_{L/R}(\kvec)|u_{L/R}(\kvec)\rangle$.
The transformation matrices $U_{L}(\kvec)$ and $U_{R}(\kvec)$ on the two hemispheres are both defined on the loop $l_2$, but are related by a transition function $O_1(\kvec)$ along the loop $l_2$:
\begin{align}
U_{R}(\kvec)=U_{L}(\kvec)O_{1}(\kvec),~~\kvec\in l_2.
    \label{eq:ULUR}
\end{align}

The third step is to relate the above-mentioned construction to the middle term $\pi_2(U(N),O(N))$ in Eq.~(\ref{eq:iso}). To this end, we focus on a specific smooth gauge for which the left and right transition matrices along the closed loop $l_2$ are $\tilde{U}_L(\kvec)=\mathbb{I}$ and $\tilde{U}_R(\kvec)=O_1(\kvec)$, respectively. 
We can choose this specific gauge for convenience and without lost of generality because all gauges that satisfy the $\TR$-constant and smooth conditions are considered equivalent. 
It follows that the right transformation matrix $\tilde{U}_R(\kvec)$ lives in the space of $U(N)$ within the internal right hemisphere, but lives in $O(N)$ on the boundary loop $l_2$. 
The map from the right hemisphere $\kvec\in R$ (including the boundary loop $l_{2}$) to the space of $\tilde{U}_R(\kvec)$ is therefore described by an element in the middle term $\pi_2(U(N),O(N))$ in Eq.~(\ref{eq:iso}). 

Under this construction, we can further interpret the right-most term in Eq.~(\ref{eq:iso}) as the first homotopy group of the transition matrix $O_1(\kvec)$ along the boundary loop $\kvec\in l_2$ of the right hemisphere, which provides a concrete way to express the 2D invariant $w_2$ in certain cases. 
For instance, for the case where $N_0=2$ and $O_1(k)$ is in the form of $e^{-i\alpha(k)\sigma_y}$, the 2D invariant can be computed as 
\begin{align}
w_2=\oint_{l_2}dk \, \pt_k \alpha(k), 
    \label{eq:w2}
\end{align}
which is the winding number of the transition function along the discontinuity loop $l_2$. 

Finally, we comment on the situations where our proposed form for $w_2$ does not hold. 
First, the construction described in Sec.~\ref{sec:2d} only holds when there is no obstruction in satisfying the $\TR$-constant and smooth gauge conditions. 
When there is no gauge under which the $\TR$-constant and smooth conditions are satisfied everywhere on the BZ, the sewing matrix $\sew_{I/II}(\kvec)$ becomes ill-defined at some point $\kvec$ in the BZ. 
Such obstruction could be detected upfront using our 1D invariants $\nu_{1a}$ and $\nu_{1c}$ [see Eq.~(\ref{eq:nuw2})] defined on 1-cells $a$ and $c$. Specifically, based on the detection of obstruction proposed in Ref.~\cite{Kruthoff2019} for cases with reflection symmetry, we conjecture that $\nu_{1a}+\nu_{1c}=1$ implies such an obstruction (see an example of 2D first-order topological superconducting phase later in Sec.~\ref{sec:TSC}). 
In this case $w_2$ is ill-defined. We note that the obstruction is intrinsic  and is not just an avoidable technical obstruction.
As seen below, the physical consequence is that the boundary supports helical Majorana edge modes protected
by the time-reversal symmetry and hence there is no way to consistently address half of the time-reversal related pair throughout the BZ. 
Second, without the obstruction, the first homotopy group of the transition function $O_1(\kvec)$, which is generally an orthogonal matrix instead of a unitary matrix, is not always easily written as an integral along $\kvec\in l_2$. 
Here a well-defined $w_2$ exists but an explicitly calculable expression for $w_2$ remains elusive. 
We therefore introduce an alternative way of expressing $w_2$ in the next subsection.   

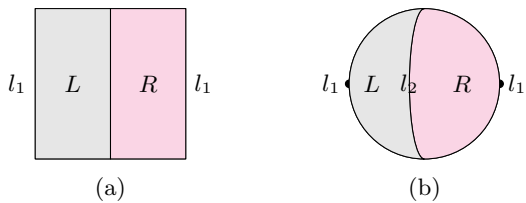
\begin{figure}
	\subcaptionbox{\label{fig:torus}}[0.4\linewidth]
		{\begin{tikzpicture}
			\draw (0,0) [black,-] --node[left] {$l_1$} ++(0,2.0);
			\draw (2.0,0) [black,-] --node[right] {$l_1$} ++(0,2.0);
			\draw (0,0) [black,-] --(2.0,0);
			\draw (0,2.0) [black,-] --(2.0,2.0);
			\draw (1.0,0) [black,-] --node {$l_2$} ++(0,2.0);
			\draw[fill=gray!20] (0,0) -- (0,2.0) -- (1.0,2.0) -- (1.0,0) -- cycle;
			\draw[fill=magenta!20] (1.0,0) -- (1.0,2.0) -- (2.0,2.0) -- (2.0,0) -- cycle;
			\node at (0.5,1.0) {$L$};
			\node at (1.5,1.0) {$R$};
		\end{tikzpicture}}
	\hspace{5mm}
	\subcaptionbox{\label{fig:sphere} }[0.4\linewidth]
		{\begin{tikzpicture}
			\tikzstyle{dot} = [draw,shape=circle,minimum size=1mm,inner sep=0pt,outer sep=0pt,fill=black];
			\draw (0,1) arc (90:270:0.2cm and 1cm);
			\draw[dashed] (0,1) arc (90:-90:0.2cm and 1cm);
			\draw (0,0) circle (1cm);
			\node[dot] at (-1cm,0) {};
			\node[dot] at (1cm,0) {};
			\draw[fill=magenta!20] (0,1) arc (90:270:0.2cm and 1cm) arc (270:450:1cm and 1cm);
			\draw[fill=gray!20] (0,1) arc (90:270:0.2cm and 1cm) arc (270:90:1cm and 1cm);
			\node[left] at (-1cm,0) {$l_1$};
			\node[right] at (1cm,0) {$l_1$};
			\node at (-0.2cm,0) {$l_2$};
			\node at (-0.7cm,0) {$L$};
			\node at (0.5cm,0) {$R$};
		\end{tikzpicture}}
	\caption{Schematics for the procedure of constructing a 2-sphere from the BZ when computing the 2D invariant $w_2$. (a) shows the BZ before the procedure, which is denoted by the rectangle. The BZ is separated into the left (L) and right (R) patches in a way that the real-gauge wavefunctions are smooth within each of the patches, but can be discontinuous across the closed loop $l_2$. We then cut the BZ open along the closed loop $l_1$ and collapse the open edges into two points to obtain the 2-sphere in (b).} \label{fig:w2}
\end{figure}

\subsection{Interpretation as Stiefel-Whitney classes}
\label{sec:SW}

Given the limitation of the $w_2$ expression we proposed in section \ref{sec:2d}, in this subsection we seek for a more generally calculable expression for $w_2$. 
Our strategy is to employ the fact that mathematically for a given real vector bundle, there are a series of $\ztwo$ invariants given by the S-W classes, which characterizes various properties of the real vector bundle. 
Specifically, the $i$-th S-W class takes value in the $i$-th cohomology class of the base space with $\ztwo$ coefficients, which indicates how similar the vector bundle is to a product vector bundle on the $i$-dimensional subspaces of the base space~\cite{VBK}. 
For instance, the first S-W class assigns the value 0 or 1 to each non-contractible loop when the bundle is orientable or non-orientable on the loop.   
Since both of the invariants $w_1$ and $w_2$ for 2D time-reversal superconductors with $C_2$ symmetry are $\ztwo$ quantities, which we know from our AHSS results in Table III, our goal is to obtain a real-vector-bundle description for the considered superconductors and examine if the S-W classes can distinguish phases that $w_1$ and $w_2$ are supposed to discern. 
This strategy of writing down topological invariants has in fact been applied to insulators with a combined symmetry $\TR\CR$ of time reversal and two-fold rotation by Refs.~\cite{Ahn2018,Ahn2019}. 
Here, we will apply a similar approach to identify our expressions for $w_1$ and $w_2$ [see Eq.~(\ref{eq:wind}) and Sec.~\ref{sec:2d}] as the first and second S-W classes of real vector bundles that describe 2D superconductors with both the time-reversal $\TR$ and two-fold rotation $\CR$ symmetries. 
Importantly, we will show that a well-known relation between the first and second S-W classes allows us to express $w_2$ in terms of $w_1$, which we know how to calculate explicitly [see Eq.~(\ref{eq:sum})]. 

We first show that $w_1$ can be identified as the first S-W class of a rank-$N_0$ real vector bundle $\VB_1$, which is defined as follows. 
The base space of $\VB_1$ is a closed loop formed by a 1-cell $j\in\{a,b,c\}$, its symmetry related 1-cell $j^\prime$, and the boundary between them (see Fig. \ref{fig:BZ}). 
For the vector space of $\VB_1$, we impose the $\TR$-constant gauge condition on the $2N_0$ occupied wavefunctions along the loop such that these wavefunctions can be split into two time-reversal-related sets of $N_0$ wavefunctions $I$ and $II$.   
We then define the vector space of $\VB_1$ as the space spanned by the occupied wavefunctions of set $I$ (or equivalently set $II$) \textit{in the real gauge}. 
The resulting vector bundle $\VB_1$ is a rank-$N_0$ real vector bundle. 

The first S-W class of the vector bundle $\VB_1$ is the orientability of $\VB_1$, which can be extracted from the set $I$ of real-gauge occupied wavefunctions as follows. 
Although it is always possible to demand the wavefunctions to be in the smooth gauge along the base space $\tilde{l}_j=l(j\cup j^\prime)$, which is a 1D loop, the wavefunctions may inevitably suffer discontinuous $\pi$-phase shifts when transformed into the real gauge. 
In such cases, 
we can detect a $\pi$-phase shift that occurs at momentum $\kvec\in\tilde{l}_j$ by the transition function matrix 
\begin{align}
O_{1,mn}(\kvec)=\langle u^\prime_{m}(\kvec_L)|u^\prime_{n}(\kvec_R)\rangle, 
\label{w1O}
\end{align}
where $|u^\prime_{n}(\kvec_{L/R})\rangle$ is the $n$-th real-gauge occupied wavefunction within the left and right neighborhoods about  $\kvec\in\tilde{l}_j$. 
Since the transition function formed by real-gauge wavefunctions is an orthogonal matrix, its determinant $\det O(\kvec)=1$ or $-1$, which corresponds to the absence or presence of a $\pi$-phase shift at $\kvec$. 
Thus, the vector bundle $\VB_1$ is said to be orientable and non-orientable if there exists even and odd numbers of momenta $\kvec$ with   
$\det O(\kvec)=-1$, respectively. 
We can therefore determine the first S-W class for each loop $\tilde{l}_j$, which is a $\ztwo$ number that labels the orientability of the vector bundle. 

We now further relate the first S-W class to our 1D invariant $w_1$ [see Eq.~(\ref{eq:wind})], which is given by the winding number of the sewing matrix $\sew_{I/II}$ of one time-reversal copy along a 1-cell loop $\tilde{l}_j$. To this end, note that the discontinuous $\pi$-phase shifts of the wavefunctions along $\tilde{l}_j$ are also embedded in the gauge transformation matrix $U$ that transforms the sewing matrix $\sew_{I/II}=UU^T$ in the smooth gauge to that in the real gauge $\sew_{I/II}=\mathbb{I}$ [see Eq.~(\ref{eq:gauge})]. Specifically, the transformation matrix across some momentum $\kvec$ at which a $\pi$-shift occurs also suffers a jump given by the transition function $O_1$ 
\begin{align}
    U(\kvec_R)=U(\kvec_L)O_{1}(\kvec), 
\end{align}
where $\kvec_{L/R}$ are momenta in left and right neighborhoods about $\kvec$. Each of the transition function $O_{1}(\kvec)$ with $\det O(\kvec)=-1$ contributes a nontrivial $\pi$ phase winding for $U(\kvec)$. This implies that the phase winding of $U(\kvec)$ along the closed loop $\tilde{l}_j$ is given by $\oint_{\tilde{l}_j}dk\partial_k\log\det U(k)=0$ or $\pi$ when the first S-W class is trivial or nontrivial, respectively. This orientability of the vector bundle $\VB_1$ is also carried by the smooth-gauge sewing matrix through the winding of $U(\kvec)$ since $\sew_{I/II}(\kvec)=U(\kvec)U(\kvec)^T$. 
Specifically, the phase winding of the sewing matrix $\sew_{I/II}(\kvec)$ along the 1D loop $\tilde{l}_j$ is given by twice of that of $U(\kvec)$. The corresponding winding number of $\sew_{I/II}$, which is just the 1D invariant $w_1$ we obtained from our AHSS result [see Eq.~(\ref{eq:wind})], is therefore given by  
\begin{align}
    w_1=\frac{1}{2\pi}\oint_{\tilde{l}_j}dk\partial_k\log\det \sew_{I/II}(k)=0~~\text{or}~~1  
\end{align}
when the first S-W class is trivial or nontrivial, respectively. 
We have thus identified our proposed 1D invariant $w_1$ as the first S-W class of the real vector bundle $\VB_1$.    

Next, we show that $w_2$ can be identified as the second S-W class of the following rank-$N_0$ real vector bundle $\VB_2$.
The base space of $\VB_2$ is the closed space formed by an orbit of 2-cells and the boundary between them, which is the entire BZ (see Fig. \ref{fig:BZ}). The vector space of $\VB_2$, similar to that of $\VB_1$, is the space spanned by the occupied wavefunctions of set $I$ (or equivalently set $II$) in the $\TR$-constant and real gauge. 
The second S-W class of a vector bundle is a $\ztwo$ quantity that  describes whether there exists a set of orthogonal basis everywhere on the BZ, and is mathematically given by the first homotopy class $\pi_1(O(N))$ of the group of orthogonal matrices. 
For the vector bundle $\VB_2$, by applying the arguments in Ref.~\cite{Ahn2019} for insulators to our superconducting case, we can identify the winding number of the orthogonal transition function $O_1(\kvec)$ (see Sec.~\ref{sec:2d}) defined on some non-contractible loop in BZ as the second S-W class of $\VB_2$. 
Since we have shown that the 2D invariant $w_2$ can also be calculated by this winding number [see the discussion near Eq.~(\ref{eq:w2})], we can identify $w_2$ as the second S-W class of $\VB_2$.  

Now that we have identified our invariants $w_1$ and $w_2$ as the first S-W class of vector
bundle $\VB_1$ and the second S-W class of $\VB_2$, respectively, we are ready to obtain a practically calculable expression for $w_2$ using the well-known Whitney sum formula 
\be
	W_n(\VB \oplus \tilde{\VB}) = \sum_{i,j \geq 0, i+j=n} W_i(\VB) \smile W_j(\tilde{\VB}),
	\label{eq:whitney}
\ee 
where $W_n$ for $n\geq1$ denotes the $n$-th S-W class with $W_0$ being the trivial cohomology class~\cite{VBK}, $\smile$ denotes the cup product, and $\VB$, $\tilde{\VB}$ are real vector bundles.
Since this formula relates higher-order to lower-order S-W classes of different vector bundles, we can apply Eq.~(\ref{eq:whitney}) to express $w_2$ in terms of $w_1$, for which we have proposed a broadly calculable expression [see Eq.~(\ref{eq:wind})]. 

To apply the Whitney sum formula to our case, we consider a given time-reversal BdG Hamiltonian $H$ with $4N_0$ bands and write $H=\bigoplus_{s=1}^{N_0} H_{sub}^s$ as the direct sum of $N_0$ four-band `sub-Hamiltonians' $H_{sub}^s$ labeled by $s$. 
We can then construct a rank-$1$ real vector bundle $\VB_s$ for each sub-Hamiltonian $H_{sub}^s$, where the base space is the BZ and the rank-$1$ vector space is spanned by one of the two time-reversal-related occupied states in the real gauge. 
The direct sum of these rank-$1$ real vector bundle $\VB_s$ leads to the vector bundle $\VB_{tot} = \bigoplus_{s=1}^{N_0} \VB_s$, for which the vector space is spanned by one of the two sets of time-reversal-related occupied states of the full BdG Hamiltonian $H$. We can now relate the S-W classes of these rank-$N_0$ and rank-1 bundles using the Whitney sum formula 
\be
	W_2(\VB_{tot}) = \sum_{(s,s^\prime)} W_1(\VB_s) \smile W_1(\VB_{s^\prime}),
	\label{eq:whitney2}
\ee
where the sum is over all unordered pairs of $(s,s^\prime)$ with $s,s^\prime=1,\cdots,N_0$. Here, we have made use of the fact that the second S-W class for a rank-$1$ bundle is always trivial such that $W_2(\VB_s)=0$ for all $s$~\cite{bundles}.  

The last step is to express the 2D invariant $w_2$ in terms of the 1D invariant $w_1$ through Eq.~(\ref{eq:whitney2}). The left-hand side of Eq.~(\ref{eq:whitney2}) corresponds to $w_2$ since we have identified $w_2$ as the second S-W class of the vector bundle $\VB_2$, and the real vector bundle $\VB_{2}$ is just $\VB_{tot}$. 
For the right-hand side of Eq.~(\ref{eq:whitney2}), the first S-W class $W_1(\VB_s)$ gives the orientabilities of $\VB_s$ along the non-contractible 1D loops in the BZ. To relate these to $w_1$'s, we focus on a bundle $\tilde{\VB}_{s,j}$ out of $\VB_s$ by restricting the base space from the full BZ to one of the non-contractible loops $\tilde{l}_j$, $j\in\{a,b,c\}$. Its orientability, previously identified with $w_{1j}$, is the same as the orientability of $\VB_s$ along $\tilde{l}_j$. 
Moreover, we only consider the two inequivalent non-contractible loops $\tilde{l}_b$ and $\tilde{l}_c$  
given that the loops $\tilde{l}_a$ and $\tilde{l}_c$ (see Fig.~\ref{fig:BZ}) belong to the same homology class. 
The right-hand side of Eq.~(\ref{eq:whitney2}) thus corresponds to the sum of all possible products of the 1D invariants $w_1$ for 4-band sub-Hamiltonians defined on orthogonal loops. 
By combining the left- and right-hand sides of the Whitney sum formula, we arrive at an expression of $w_2$ in terms of $w_1$  
\be
	w_2 = \sum_{(s,s^\prime)} ( w_{1b}^s w_{1c}^{s^\prime} + w_{1c}^s w_{1b}^{s^\prime} ),
	\label{eq:sum}
\ee
where $w_{1j}^s$ can be calculated by Eq.~(\ref{eq:wind}) for a sub-Hamiltonian $H_{sub}^s$ defined on the 1-cell loop $\tilde{l}_j$. In the above we focus on the case without obstruction to the $\TR$-constant and smooth gauge conditions, where $w_2$ and $\VB_2$ are well-defined. Note that there is no nice decomposition for the $w_2$ invariant in contrast to the case of 1-D invariants (Eq.~\ref{eq:nuw2}). It is thus necessary to compute the invariant $w_2$ for the total Hamiltonian, including the reference Hamiltonian when there is a momentum-dependence in the $\mathcal{TR}$ symmetry operator. The invariant $w_2$ computed in this way is by definition the invariant $\nu_2$ in Eq.~\ref{eq:nuw}.

\section{Bulk-boundary correspondence}
\label{sec:real}

Equipped with calculable expressions for the 1D and 2D invariants $w_1$ and $w_2$, we now show that instead of just diagnosing the bulk topology, these bulk invariants are also capable of diagnosing the type of boundary Majorana modes. To establish this bulk-boundary correspondence for our case study, our strategy is to turn to classification analyses in the real space, where the boundary modes for each phase can be most naturally studied. In this section, we first study the real-space classification as well as the Majorana boundary type of each phase for 2D time-reversal superconductors with $C_2$ symmetry. We then study our invariants $\nu_{1j}$, $j=a,b,c$ and $\nu_2$ for each of these phases, whose Majorana boundary type is known, to understand how to predict boundary signatures from the momentum-space invariants. This is the key step in our protocol, which combines the momentum-space and real-space classification results to establish the bulk-boundary correspondence for our case study.

\subsection{Real-space classification}
\label{sec:real_class}

The real-space classification method we adopt is the topological crystal approach~\cite{Huang2017,Song2017,shiozaki2018generalized,Song2019,Song2020,Song2020real,Huang20204d,huang2021effective}, which is a general real-space treatment for topological crystalline phases, and is dual to the momentum-space classification method AHSS we used in Sec.~\ref{sec:AHSS}. The main idea of this method is that any $d$-dimensional topological crystalline state is adiabatically connected to a real-space stacking of `building blocks', which are $d_{b}$-dimensional topological states with $d_{b} \leq d$ that do not exhibit crystalline symmetries.  
From this building block picture, one can readily determine the boundary signature of each of the topological crystalline superconducting phases with $C_2$ symmetry. We will show in the following that all the non-interacting resultant states built by the real-space building blocks for this symmetry class can be characterized by our proposed invariants $w_1$ and $w_2$.  

The first step of the topological crystal approach is to perform the $C_{2}$ symmetric cell decomposition in the real space. Fig.~\ref{fig:unit_cell} is a decomposition of the unit cell that respects the required symmetries in the real space. The high-symmetry points are marked with their coordinates. The independent 1D cells are $e^{(1)}_1$, $e^{(1)}_2$, and $e^{(1)}_3$; the independent 2D cell is $e^{(2)}$. All the other cells are related to them by symmetries. 

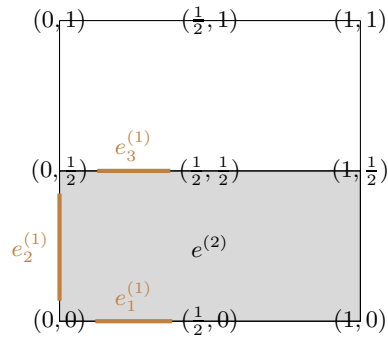
\begin{figure}[h]
	\centerline{
	\begin{tikzpicture}
		\tikzstyle{zeroc} = []
		\tikzstyle{onec} = [black,-]
		\draw[fill=gray!30] (-2.0,-2.0) -- (2.0,-2.0) -- (2.0,0) -- (-2.0,0) -- cycle;
		\node[zeroc] (q11) at (0,0) {$(\frac{1}{2},\frac{1}{2})$};
		\node[zeroc] (q21) at (2.0,0) {$(1,\frac{1}{2})$};
		\node[zeroc] (q01) at (-2.0,0) {$(0,\frac{1}{2})$};
		\node[zeroc] (q12) at (0,2.0) {$(\frac{1}{2},1)$};
		\node[zeroc] (q10) at (0,-2.0) {$(\frac{1}{2},0)$};
		\node[zeroc] (q20) at (2.0,-2.0) {$(1,0)$};
		\node[zeroc] (q02) at (-2.0,2.0) {$(0,1)$};
		\node[zeroc] (q22) at (2.0,2.0) {$(1,1)$};
		\node[zeroc] (q00) at (-2.0,-2.0) {$(0,0)$};
		\node (c1) at (0,-1.0) {$e^{(2)}$};
		\draw (-2.0,0) [onec] -- (2.0,0);
		\draw (-2.0,2.0) [onec] --(2.0,2.0);
		\draw (-2.0,-2.0) [onec] --(2.0,-2.0);
		\draw (-2.0,-2.0) [onec] --(-2.0,2.0);
		\draw (2.0,-2.0) [onec] --(2.0,2.0);
		\draw (q00) [brown,ultra thick,-] -- node[above] {$e^{(1)}_1$} ++(q10);
		\draw (q00) [brown,ultra thick,-] -- node[left] {$e^{(1)}_2$} ++(q01);
		\draw (q01) [brown,ultra thick,-] -- node[above] {$e^{(1)}_3$} ++(q11);
	\end{tikzpicture}
	}
	\caption{The symmetry-preserving cell decomposition for a unit cell in the real space. The coordinates $(x,y)$ denote the 0-cells, the brown lines $e^{(1)}_i$, $i=1,2,3$ denote the three independent 1-cells, and the shaded area in grey $e^{(2)}$ denotes the independent 2-cell.}
\label{fig:unit_cell}
\end{figure}

The second step is to identify the 0D, 1D, and 2D building blocks that live on different dimensional cells. For our systems, there are no 0D building blocks since the classification of the effective 0D BdG Hamiltonian is trivial. This can be seen from the following. The two irreps of $C_2$ rotation labeled by $\CR=\pm i$ are mapped to each other by $\TR$ or $\PH$. At any high symmetry point (0-cells), the emergent AZ class for either irrep is therefore AIII since the chiral symmetry $\PH\TR$ maps one irrep to itself. Since the classification for 0D class-AIII BdG Hamiltonians is trivial, there are no nontrivial 0D building blocks in our topological crystal analysis. 
We nonetheless have nontrivial 1D and 2D building blocks. 
Since the 1-cells and 2-cells (i.e. any point in a unit cell besides the high-symmetry points) exhibit emergent time-reversal and particle-hole symmetries, there exist a 1D building block of a time-reversal-invariant Kitaev chain (spinful Kitaev chain) and a 2D building block of a 2D time-reversal-invariant topological superconductor (TSC) with helical edge states. 

By stacking these nontrivial building blocks into 2D superconducting states in different symmetry-allowed ways, we can obtain the representative states for all topologically distinct phases in the considered symmetry class. 
These representative states built by building blocks, which are dubbed the building-block states, can be generated from the following four fundamental block constructions: The first three are 1D-block states built by placing a spinful Kitaev chain on one of the 1-cells $e^{(1)}_1$, $e^{(1)}_2$, and $e^{(1)}_3$ in every unit cell. The last one is a 2D-block state built by placing a 2D TSC on the 2-cell $e^{(2)}$ in every unit cell. 
The rest of the representative states can then be generated by the direct sums of these 1D- and 2D-block states.

Based on the topological crystal classification, the classifying group $\tilde{K}$ of the 2D time-reversal superconductors with $C_2$ symmetry is described by the short exact sequence 
\be
	1 \rightarrow \ztwo^3 \rightarrow \tilde{K} \rightarrow \ztwo \rightarrow 1.
	\label{eq:K_ES_rs}
\ee
Here, the entry $\ztwo^3$ is generated by the 1D block states with a spinful Kitaev chain placed on each of the three 1-cells $e^{(1)}_i$, $i=1,2,3$, and the entry $\ztwo$ is generated by the 2D block state with a 2D TSC placed on the 2-cell $e^{(2)}$. The full classification group $\tilde{K}$ is given by the group extension of $\ztwo$ by $\ztwo^3$. 
By examining the resultant state from stacking two copies of 2D TSC's (see Appendix~\ref{app:edge}), we show that the group extension is trivial so that the classifying group $\tilde{K}=\ztwo^4$. 

Now we comment on the relation between the real-space and the momentum space classification. The AHSS calculation leads to the subgroups of $K$ that obey a short exact sequence [see Eq.~(\ref{eq:K_ES})], but does not reveal the full group structure of $K$. To determine $K$, one needs to solve the group extension problem for Eq.~(\ref{eq:K_ES}), which can be done with the help of the block-state picture (see Appendix~\ref{app:group}). We find that $K = \mathbb{Z}_{2}^{4}$, which matches with the real space classification $\tilde{K} = \mathbb{Z}_{2}^{4}$. However, the group multiplication rule of $K$ is ambiguous. Finally, we note that the information about the group extension is not necessary for the purpose of finding momentum-space invariants.

\subsection{Boundary signatures of building-block states}
\label{sec:block_boundary}

The boundary signatures of these building block states can be naturally obtained from their block constructions.
For each building-block state, we describe the protecting symmetry and the Majorana boundary signature separately using the following terminology. Depending on whether the protecting symmetry of the state is the translational symmetry or one of the time-reversal and $C_2$ symmetries, the state is said to belong to a weak or strong phase, respectively. 
On the other hand, we define the order of a state with symmetric boundary terminations to be the lowest co-dimension of boundaries with gapless Majorana boundary modes. 
Following our definition, the order of a state directly indicates the dimension of Majorana boundary modes that can be probed in experiments. 
For example, a state that supports gapless Majorana modes on partial or all edges is a first-order state, regardless of the protecting symmetry. 
A state that supports Majorana zero modes at two $C_2$-related corners is a second-order state. 
Finally, a state built by stacking a first-order and a second-order states is considered first order in our definition. In this case, we expect only Majorana edge modes are visible under experimental probes since the corner modes are embedded in the edge modes. 

We now discuss the protecting symmetries and the boundary signatures of a few important building block states. First, the 2D-block state built by placing 2D TSCs on 2-cells $e^{(2)}$ is clearly a first-order strong state that is protected by the time-reversal symmetry and hosts helical Majorana edge modes.
Second, a second-order strong state with Majorana corner modes can be obtained by building a 1D-block state with one spinful Kitaev chain on each of the 1-cells $e^{(1)}_1$ and $e^{(1)}_3$~\cite{Vu2020,Huang2021}. This can be understood as follows. 
When we break the $C_2$-rotational symmetry, all the 1D blocks on $e^{(1)}_3$ can move toward the 1D blocks on $e^{(1)}_1$ and annihilate in pairs without breaking the translational symmetry. In this case, no boundary signature survives and the state is trivialized by the breaking of $C_2$ alone.  
In contrast, when we break the translation but preserve the $C_2$ rotational symmetry, most 1D blocks can still annihilate in pairs in a $C_2$-symmetric way but the one on $e^{(1)}_1$ that threads through the $C_{2z}$-rotational axis will survive. Due to this single 1D block that survives due to $C_2$, the resulting state supports two $C_2$-related 0D Majorana zero modes and is therefore a second-order strong phase protected by the $C_2$ symmetry. 

We can further achieve a weak state by a 1D-block state built with a spinful Kitaev chain on the 1-cell $e^{(1)}_3$. This state is purely weak because when the translational symmetry is broken, the Kitaev chains in different unit cells can move to annihilate each other completely such that no boundary signature is left. Under our definition, this weak state is first order since we expect Majorana bands along $y$-direction edges. Similarly, the first-order weak state with Majorana bands along $x$-edges can be constructed by placing one spinful Kitaev chain on each of the three 1-cells. 

Finally, the 1D-block state built by placing a spinful Kitaev chain on $e^{(1)}_1$ or $e^{(1)}_2$, which are the 1-cells that thread through the $C_{2z}$-rotational axis, can be viewed as the combination of a first-order weak state and a second-order strong state.
This is because when we break the $C_2$-rotational symmetry, all 1D blocks (the spinful Kitaev chains) can still survive under the translational protection such that the state supports Majorana bands along $y$- or $x$-edges. 
In contrast, when we break the translation but preserve the $C_2$ rotational symmetry, all 1D blocks can annihilate in pairs in a $C_2$-symmetric way except the spinful Kitaev chain that threads through the $C_{2z}$-rotational axis. Due to this single 1D block, the resulting state supports two $C_2$-related 0D Majorana zero modes and is therefore a second-order strong phase protected by the $C_2$ symmetry. 
As a combination of the above two states, this 1D block state is considered a mixture of weak and strong states in terms of the protecting symmetries, and is considered first order in terms of the boundary signature under our definition. 

We emphasize that these building-block states serve as the representatives of the whole phase---any other state is adiabatically connected to a building-block state if they are in the same phase (see Sec.~\ref{sec:real_class}.). We therefore expect that the same Majorana boundary signature found in a representative building-block state is shared by the entire phase to which it belongs.  
In the following, we will construct minimal models for the four fundamental 1D- and 2D-block states and explicitly compute our proposed topological invariants $w_1$ and $w_2$ [defined in Eq.~(\ref{eq:wind}) and Eq.~(\ref{eq:sum})]. 
Through such calculations, we will be able to establish how $w_1$ and $w_2$ are related to the Majorana boundary signature of each of the topologically distinct phases.

\subsection{1D-block states}
\label{sec:KC}

We start from the 1D-block states. 
First, we focus on the state formed by stacking spinful Kitaev chains~\cite{Shiozaki2019} on the 1-cells $e^{(1)}_1$ only. According to our topological crystal analysis in Sec.~\ref{sec:block_boundary}, this state belongs to a first-order phase with Majorana bands along the $y$-directional edges. 
To explicitly compute our proposed invariants $w_1$ and $w_2$, we consider the following minimal BdG model 
\be
	H_{KC}(\kvec) = -(\cos{k_x}+m) \tau_z\sigma_0 + \sin{k_x} \tau_x\sigma_z,  
	\label{eq:KC} 
\ee
where $\abs{m}<1$, and $\tau_i$ and $\sigma_i$ are $2\times2$ Pauli matrices in the particle-hole and spin spaces, respectively. We denote the vector bundle on which this model $H_{KC}$ acts by $\VB_{e^{(1)}_1}$, and the symmetry operators respecting the commutation relations Eq.~(\ref{eq:sym_ham}) and Eq.~(\ref{eq:sym}) are given by 
\be
	\TR = i\sigma_y \cc, \, \PH = \tau_x \cc, \, \CR = i\tau_z \sigma_z,
	\label{eq:KC_sym}
\ee
where $\cc$ is complex conjugation. 

To compute the proposed topological invariants, the first step is to write down the wavefunctions explicitly in the basis of the eigenstates $|\pm\pm\rangle$ of both $\tau_z$ and $\sigma_z$, where the former and latter $\pm$ denote the corresponding $\tau_z$ and $\sigma_z$ eigenvalues $\pm1$, respectively. 
In the basis of $\{++,+-,-+,--\}$, the two real-gauge occupied states of energy $-\epsilon_0$ are
\begin{align}
	\psi_1 (\kvec) = \frac{1}{2\sqrt{\epsilon_0(k_x)}} \begin{pmatrix}
	i\sqrt{\cos{k_x}+m+\epsilon_0} \\
	\sqrt{\cos{k_x}+m+\epsilon_0} \\
	-i\frac{\sin{k_x}}{\sqrt{\cos{k_x}+m+\epsilon_0}} \\
	\frac{\sin{k_x}}{\sqrt{\cos{k_x}+m+\epsilon_0}}
	\end{pmatrix}, \nonumber \\
	\psi_2 (\kvec) = \frac{1}{2\sqrt{\epsilon_0(k_x)}} \begin{pmatrix}
	\sqrt{\cos{k_x}+m+\epsilon_0} \\
	i\sqrt{\cos{k_x}+m+\epsilon_0} \\
	-\frac{\sin{k_x}}{\sqrt{\cos{k_x}+m+\epsilon_0}} \\
	i\frac{\sin{k_x}}{\sqrt{\cos{k_x}+m+\epsilon_0}}
	\end{pmatrix},
	\label{eq:KC_real}
\end{align}
where $\epsilon_0(k_x)=\sqrt{m^2+2m\cos{k_x}+1}$. Note that these two occupied states are Kramers pairs such that $\TR \psi_1(\kvec) \propto \psi_2(-\kvec)$, and they are not continuous across $k_x=\pi$. 

The second step is to write down the reference Hamiltonian $H_0$. By taking $m$ in $H_{KC}$ to the negative infinity, we find 
\be
	H_0 = \tau_z\sigma_0,
	\label{eq:H0}
\ee
which acts on $\VB_{e^{(1)}_1}$ and the symmetry operators have the same forms as in Eq.~(\ref{eq:KC_sym}). The occupied states of $H_0$ are two degenerate $\kvec$-independent flat bands with a $\tau_z\sigma_0$-eigenvalue of $-1$. 
Due to the $\kvec$-independence, both the 1D and 2D invariants vanish for the reference Hamiltonian $H_0$ such that 
\begin{align}
	& \nu_{1j} (\VB_{e^{(1)}_1}, H_{KC}, H_0)=w_{1j} (\VB_{e^{(1)}_1}, H_{KC}), \, j=a, b, c, \nonumber \\
	& \nu_2 (\VB_{e^{(1)}_1}, H_{KC}, H_0)=w_2 (\VB_{e^{(1)}_1}, H_{KC}), 
\end{align}
where $j=a,b,c$ labels the 1-cells in the momentum space (see Fig. \ref{fig:BZ}). 
We can therefore directly calculate $w_{1j}$ and $w_2$ for $H_{KC}$ without worrying the contribution from $H_0$. 

The last step is to compute $w_{1j}$ following Eq.~(\ref{eq:w1}) and $w_2$ as the second S-W class.
For the 1-cell $j=a$, we restrict the real-gauge wavefunctions $\psi_1(\kvec)$ and $\psi_2(\kvec)$ on the closed loop $l(a\cup a^\prime)$ in the momentum space (see Fig. \ref{fig:BZ}).  
We then perform a gauge transformation $U=e^{-i\frac{k_x}{2}} \iden_{2\times2}$ from the real gauge to a smooth and $\TR$-constant gauge, where the identity  $\iden_{2\times2}$ acts on the eigen-basis of $\psi_1$ and $\psi_2$. In this gauge, the sewing matrix becomes $\sew=e^{ik_x} \iden_{2\times2}$ such that the 1D invariant on 1-cell $j=a$ is clearly $w_{1a} = 1$ (see Eq.~\ref{eq:w1}). Similar treatment for the other 1-cells $j=b, c$ shows that $w_{1b}=0, \, w_{1c}=1$. 
For the 2D invariant $w_2$, we interpret $w_2$ as the second S-W class of the rank-1 vector bundle $\tilde{\VB}$ formed by either $\psi_1$ or $\psi_2$. This can be done because the 1D real vector space spanned by one of the two real-gauge wavefunctions $\psi_{1/2}(\kvec)$ is well-defined in the entire BZ.  
Since the second S-W class for a rank-1 vector bundle is always trivial, we find $w_2=0$. 
To summarize, we find that the 2D superconducting state formed by stacking one spinful Kitaev chain per 1-cell $e_1^{(1)}$ in the real space is characterized by the following 1D and 2D topological invariants defined in the momentum space 
\begin{align}
    &\nu_{1a}=\nu_{1c}=1, ~~\nu_{1b}=0,~~\nonumber\\
    &\nu_2=0.   
    \label{eq:nue1}
\end{align}
Correspondingly, in the real space this superconducting state supports Majorana bands along the $y$-directional edges (with undetectable Majorana corner modes buried in the Majorana bands).

Similar to the above case but with $k_x$ and $k_y$ interchanged, the 1D-block state formed by stacking spinful Kitaev chains on $e^{(1)}_2$ supports Majorana bands along the $x$-directional open edges (also with embedded Majorana corner modes). With a similar calculation, we find that this state is characterized by 1D and 2D topological invariants $\nu_{1a}=\nu_{1c}=0, \, \nu_{1b}=1, \, \nu_2=0$. 

Next, we consider the 1D-block state formed by stacking spinful Kitaev chains on the real-space 1-cells $e^{(1)}_3$, where we denote the corresponding vector bundle as  $\VB_{e^{(1)}_3}$. Importantly, while the system and the reference Hamiltonians are also given by $H_{KC}$ in Eq.~(\ref{eq:KC}) and $H_0$ in Eq.~(\ref{eq:H0}), respectively, the symmetry operator of $C_2$ rotation now takes a different form from Eq.~(\ref{eq:KC_sym}) since none of the 1-cells $e_3^{(1)}$ encounters the $C_{2z}$ rotational axis. Specifically, the symmetry operators are given by
\be
	\TR = i\sigma_y \cc, \, \PH = \tau_x \cc, \, \CR = ie^{ik_y}\tau_z \sigma_z, 
	\label{eq:KC3_sym}
\ee
where $\CR$ acquires extra momentum dependence. This extra factor $e^{ik_y}$ arises from the fact that upon rotation, the position $y=\frac{1}{2}$ in any unit cell is mapped to the position $y=-\frac{1}{2}$ in another unit cell, which differs from the original position by an integer multiple of lattice vectors in the $-y$ direction~\cite{Shiozaki2017,Shiozaki2019}. This 1D-block state, which is topologically distinct from those formed by stacking Kitaev chains on $e^{(1)}_1$ or $e^{(1)}_2$, is a weak phase protected by the translation symmetry alone and supports Majorana bands along the $y$-directional edges without any embedded Majorana corner modes (see Sec.~\ref{sec:block_boundary}.). 

For the 1D invariants, note that the phase of the sewing matrix $\sew_{I/II}(\kvec)$ along a 1-cell loop $\tilde{l}_j$ acquires nontrivial winding from the explicit $k_y$-dependence of the combined symmetry operator  $\TR\CR=ie^{-ik_y}\tau_z \sigma_x \cc$. 
In such a case, the 1D topological invariants $w_{1j}$ of not only the system Hamiltonian $H_{KC}$, but also the reference Hamiltonian $H_0$ can become nontrivial. 
Thus, to correctly obtain the 1D invariant $\nu_{1j}$ that characterizes the weak phase this 1D-block state belongs to, we need to add the contributions from both Hamiltonians, as we discussed in Sec.~\ref{sec:refH} and Eq.~(\ref{eq:nuw3}).   
Specifically, for a momentum-space 1-cell $j=a,b,c$, we have 
\begin{align}
	& \nu_{1j} (\VB_{e^{(1)}_3}, H_{KC}, H_0) = w_{1j}(\VB_{e^{(1)}_3}, H_{KC}) + w_{1j}(\VB_{e^{(1)}_3}, H_0),
\end{align}
where the invariants for the reference Hamiltonian are $w_{1a}(\VB_{e^{(1)}_3}, H_{KC})=w_{1c}(\VB_{e^{(1)}_3}, H_{KC})=0$, $w_{1b}(\VB_{e^{(1)}_3}, H_{KC})=1$, whereas the invariants for the system Hamiltonian are $w_{1j}(\VB_{e^{(1)}_3}, H_{0})=1$ for all $j=a,b,c$. We thus arrive at 1D invariants $\nu_{1a}=\nu_{1c}=1$ and $\nu_{1b}=0$. 

For the 2D invariant, same as the previous 1D-block states, we can again interpret $w_2$ as the second S-W class of the rank-1 vector bundle $\tilde{\VB}$ formed by either $\psi_1$ or $\psi_2$. 
The only difference is that when applying the Whitney sum formula in Eq.~(\ref{eq:whitney2}), the total Hamiltonian for the second S-W class on the left-hand side is a direct sum $H_{KC} \oplus H_0$ of the system and reference Hamiltonians. 
Therefore, in the resulting $w_2$-$w_1$ relation in Eq.~(\ref{eq:sum}), we need to take the two different sub-Hamiltonian indices $s$ and $s^\prime$ to be $H_{KC}$ and $H_0$. 
From the 1D invariants we find for the system and reference Hamiltonians above, it is clear that $w_2(\VB_{e^{(1)}_3}\oplus\VB_{e^{(1)}_3}, H_{KC}\oplus H_0)=1$. 
The 2D invariant for the phase is thus given by 
\begin{align}
	& \nu_{2} (\VB_{e^{(1)}_3}, H_{KC}, H_0) = w_{2}(\VB_{e^{(1)}_3}\oplus\VB_{e^{(1)}_3}, H_{KC}\oplus H_0)=1, 
\end{align}
where we have chosen $\VB_c=\VB_{e^{(1)}_3}$ and $H_c=H_0$ in Eq.~(\ref{eq:nuw}).  
.In Appendix.~\ref{app:trans_wind} we provide another equivalent calculation of $\nu_2$ using the alternative expression we propose for $w_2$ in Sec.~\ref{sec:2d}, where $w_2$ is directly calculated from the winding number of the transition function. 

To summarize, we find that the 1D-block state formed by stacking one spinful Kitaev chain per 1-cell $e_3^{(1)}$ in the real space is characterized by the momentum-space topological invariants
\begin{align}
    &\nu_{1a}=\nu_{1c}=1, ~~\nu_{1b}=0,~~\nonumber\\
    &\nu_2=1.   
\end{align}
This superconducting state shares the same 1D invariants $\nu_{1j}$ with the state of stacked chains on $e_{1}^{(1)}$ [see Eq.~(\ref{eq:nue1})], but has a different 2D invariant $\nu_2$. 
In terms of real-space boundary signatures, this state supports Majorana bands along the $y$-directional edges \textit{without} any embedded Majorana corner modes. 

Finally, we move to the 1D-block state that belongs to a second-order strong phase with $C_2$-protected Majorana corner modes. Such a state can be built by placing spinful Kitaev chains on both of the real-space 1-cells $e_1^{(1)}$ and $e_3^{(1)}$. 
In this case, the system and reference Hamiltonians $H_{HO}$ and $H_{0}^\prime$ are given by 
\begin{align}
	&H_{HO}(\kvec) = H_{KC}(\kvec) \oplus H_{KC}(\kvec),~~\nonumber\\
	&H_0^\prime = H_0 \oplus H_0, 
	\label{eq:HO_H}
\end{align}
where both act on the same vector bundle $\VB_{HO}=\VB_{e^{(1)}_1} \oplus \VB_{e^{(1)}_3}$. 
On this vector bundle $\VB_{HO}$, the relevant symmetry operators have the form 
\be
	\TR = i\sigma_y \cc, \, \PH = \tau_x \cc, \, \CR=i\tau_z \sigma_z \oplus i\tau_z \sigma_z e^{ik_y}.
	\label{eq:HO_sym}
\ee

We can obtain the 1D invariants from those we found for the 1D-block states of stacked Kitaev chains on $e_1^{(1)}$ and $e_3^{(1)}$ since the 1D invariant $w_{1j}$ for a direct sum of multiple Hamiltonians is the sum of their individual 1D invariants [see the expression of $w_{1j}$ in Eq.~(\ref{eq:w1})]. 
Specifically, since we have
\begin{align}
  &(w_{1a},w_{1b},w_{1c})=(1,0,1)~~~~\text{for}~ w_{1j}(\VB_{e_1^{(1)}},H_{KC}),\nonumber\\
  &(w_{1a},w_{1b},w_{1c})=(0,0,0)~~~~\text{for}~ w_{1j}(\VB_{e_1^{(1)}},H_{0}),\nonumber\\
  &(w_{1a},w_{1b},w_{1c})=(1,1,1)~~~~\text{for}~ w_{1j}(\VB_{e_3^{(1)}},H_{KC}),\nonumber\\
  &(w_{1a},w_{1b},w_{1c})=(0,1,0)~~~~\text{for}~ w_{1j}(\VB_{e_3^{(1)}},H_{0}) 
  \label{eq:w1_1DH}
\end{align}
for the system and reference Hamiltonians of the $e_1^{(1)}$-state and $e_3^{(1)}$-state, we find that the 1D invariants for the higher-order Hamiltonian $H_{HO}$ and the reference Hamiltonian $H_0^\prime$ in Eq.~(\ref{eq:HO_H}) are given by 
\begin{align}
  &(w_{1a},w_{1b},w_{1c})=(0,1,0)~~~~\text{for}~ w_{1j}(\VB_{HO},H_{HO}),\nonumber\\
  &(w_{1a},w_{1b},w_{1c})=(0,1,0)~~~~\text{for}~ w_{1j}(\VB_{HO},H_{0}^\prime). 
  \label{eq:w1_1DH0}
\end{align}
The final $\ztwo$ 1D invariants $\nu_{1j}$ for the second-order phase are therefore trivial for all three momentum-space 1-cells $j=a,b,c$ 
\begin{align}
	&\nu_{1j}(\VB_{HO}, H_{HO}, H_0^\prime)\nonumber \\ 
	& = w_{1j}(\VB_{HO},H_{HO}) + w_{1j}(\VB_{HO},H_0^\prime)
	= 0,  
	\label{eq:HOnu1}
\end{align}
where we have employed Eq.~(\ref{eq:nuw3}) to incorporate the contribution from the reference Hamiltonian. Thus, the 1D invariants $\nu_{1j}$ alone cannot determine whether a superconducting state is in the trivial or second-order phase, and it is necessary to consider the 2D invariant $\nu_{2}$.

To obtain the 2D invariant for the second-order phase, we can make use of the $w_2$-$w_1$ relation in Eq.~(\ref{eq:sum}) to obtain $w_2$ from the 1D invariants $w_{1j}$'s of the $e_1^{(1)}$- and $e_3^{(1)}$-states listed in Eqs.~(\ref{eq:w1_1DH}) and (\ref{eq:w1_1DH0}).  
Specifically, first we account for the nontrivial contribution from the reference Hamiltonian by choosing 
\begin{equation}
    \VB_c=\VB_{e_3^{(1)}} \ \text{and} \ H_c=H_0.
\end{equation}
in Eq.~(\ref{eq:nuw}) to ensure that $w_2(\VB_{HO}\oplus\VB_c,H_0^\prime\oplus H_c)=0$.  
Then the 2D invariant $\nu_{2}$ for the second-order phase is directly given by the 2D invariant $w_2(\VB_{HO}\oplus\VB_c,H_{HO}\oplus H_c)$ of the modified Hamiltonian $H_{HO}\oplus H_c$. Now we can apply the $w_2$-$w_1$ relation Eq.~(\ref{eq:sum}) to calculate $w_2(\VB_{HO}\oplus\VB_c,H_{HO}\oplus H_c)$, where the summation on the right-hand side of Eq.~(\ref{eq:sum}) is over all  unordered pairs among the sub-Hamiltonians $H_{KC}$, $H_{KC}$, and $H_c$ defined on the bundles $\VB_{e_1^{(1)}}$, $\VB_{e_3^{(1)}}$, and $\VB_{e_3^{(1)}}$, respectively. Written more explicitly, the 2D invariant for the higher-order phase is given by
\begin{widetext}
\begin{align}
	&~\nu_2(\VB_{HO}, H_{HO}, H_0^\prime) = w_2(\VB_{HO}\oplus \VB_{e^{(1)}_3}, H_{HO} \oplus H_c)  \nonumber \\
	=&~w_{1b}(\VB_{e_1^{(1)}},H_{KC})w_{1c}(\VB_{e_3^{(1)}},H_{KC})+w_{1b}(\VB_{e_3^{(1)}},H_{KC})w_{1c}(\VB_{e_3^{(1)}},H_{0})
	+w_{1b}(\VB_{e_1^{(1)}},H_{KC})w_{1c}(\VB_{e_3^{(1)}},H_{0})\nonumber \\
	+&~w_{1c}(\VB_{e_1^{(1)}},H_{KC})w_{1b}(\VB_{e_3^{(1)}},H_{KC})+w_{1c}(\VB_{e_3^{(1)}},H_{KC})w_{1b}(\VB_{e_3^{(1)}},H_{0})+w_{1c}(\VB_{e_1^{(1)}},H_{KC})w_{1b}(\VB_{e_3^{(1)}},H_{0})\nonumber \\
	=&~0\times1+1\times0+0\times0+1\times1+1\times1+1\times1 ~~~\text{mod}~ 2\nonumber \\
	=&~1.
	\label{eq:HOnu2}
\end{align}
\end{widetext}

To summarize, for the higher-order phase with $C_2$-protected Majorana corner modes, we find the 1D and 2D invariants to be $\nu_{1j}=0$, $j=a,b,c$ and $\nu_2=1$ [see Eqs.~(\ref{eq:HOnu1}) and (\ref{eq:HOnu2})].

\subsection{2D-block state}
\label{sec:TSC}

There is only one 2D-block state, which belongs to the time-reversal-protected topological superconducting phase with helical Majorana edge modes (2D TSC). To compute our 1D and 2D invariants $w_1$ and $w_2$, we consider the following minimal BdG model~\cite{Shiozaki2019}
\begin{align}
	& H_{TSC}(\kvec) = -(\cos{k_x}+\cos{k_y}-m) \tau_z\sigma_0 \nonumber \\
	& ~~~~~~~~~~~~~~~ + \sin{k_x} \tau_x\sigma_z + \sin{k_y} \tau_y\sigma_0, 
	\label{eq:TSC_H}
\end{align}
where we set $\abs{m}<2$, and $\tau$ and $\sigma$ represent the Pauli matrices for the particle-hole and spin spaces, respectively.
We denote the vector bundle on which the Hamiltonian $H_{TSC}$ acts by $\VB_{e^{(2)}}$, and the symmetry operators respecting the commutation relations are given by
\begin{align}
	& \TR = i\sigma_y \cc, \, \PH = \tau_x \cc, \, \CR = i\tau_z \sigma_z, 
	\label{eq:TSC_sym}
\end{align}
where $\cc$ is complex conjugation.
The real-gauge wavefunctions of the two occupied states are given by 
\begin{align}
	\phi_1 (\kvec) = \frac{1}{\sqrt{2}} \begin{pmatrix}
	(\frac{\eta_0+\cos{k_x}+\cos{k_y}-m}{2\eta_0(\sin^2{k_x}+\sin^2{k_y})})^{\frac{1}{2}} (i\sin{k_x} + \sin{k_y}) \\
	(\frac{\eta_0+\cos{k_x}+\cos{k_y}-m}{2\eta_0(\sin^2{k_x}+\sin^2{k_y})})^{\frac{1}{2}} (\sin{k_x} + i\sin{k_y}) \\
	-i(\frac{\sin^2{k_x}+\sin^2{k_y}}{2\eta (\eta_0+\cos{k_x}+\cos{k_y}-m)})^{\frac{1}{2}} \\
	(\frac{\sin^2{k_x}+\sin^2{k_y}}{2\eta (\eta_0+\cos{k_x}+\cos{k_y}-m)})^{\frac{1}{2}}
	\end{pmatrix}, \nonumber \\
	\phi_2 (\kvec) = \frac{1}{\sqrt{2}} \begin{pmatrix}
	(\frac{\eta_0+\cos{k_x}+\cos{k_y}-m}{2\eta_0(\sin^2{k_x}+\sin^2{k_y})})^{\frac{1}{2}} (\sin{k_x} - \sin{k_y}) \\
	(\frac{\eta_0+\cos{k_x}+\cos{k_y}-m}{2\eta_0(\sin^2{k_x}+\sin^2{k_y})})^{\frac{1}{2}} (i\sin{k_x} - \sin{k_y}) \\
	-(\frac{\sin^2{k_x}+\sin^2{k_y}}{2\eta (\eta_0+\cos{k_x}+\cos{k_y}-m)})^{\frac{1}{2}} \\
	i(\frac{\sin^2{k_x}+\sin^2{k_y}}{2\eta (\eta_0+\cos{k_x}+\cos{k_y}-m)})^{\frac{1}{2}}
	\end{pmatrix},
	\label{eq:TSC_real}
\end{align}
in the eigenbases $\{++,+-,-+,--\}$ of $\tau_z$ and $\sigma_z$, where each of the former and latter signs $\pm$ denote the eigenvalues $\pm 1$ of $\tau_z$ and $\sigma_z$, respectively. 
Here, the momentum-dependent parameter $\eta_0$ is given by 
$\eta_0(\kvec)=\sqrt{(\cos{k_x}+\cos{k_y}-m)^2+\sin^2{k_x}+\sin^2{k_y}}$. 
We point out that although both $-2<m<0$ and $0<m<2$ regimes contain the 2D TSC phase and support helical Majorana edge modes when placed next to vacuum, they are topologically distinct in the sense that when placed next to each other there will be gapless Majorana edge modes localized on the boundary. In fact, the two regimes differ by a weak phase (see Appendix.~\ref{app:edge}), but one cannot tell them apart from their boundary modes against the vacuum because the edge signature of their difference is hidden in the bands of the helical states.
In the following, we will study the invariants for both regimes. 

To compute the invariants $\nu_{1j}$ and $\nu_2$ for this 2D TSC phase [see Eq.~(\ref{eq:nuw})], we first obtain the reference Hamiltonian by taking $m$ in $H_{TSC}$ to the positive infinity. We find that the reference Hamiltonian is $H_0=\tau_z\sigma_0$. 
Since the symmetry operators have no explicit $\kvec$-dependence [see Eq.~(\ref{eq:TSC_sym})], the reference Hamiltonian $H_0$ does not contribute to the invariants $w_{1j}$ and $w_2$ defined for Hamiltonians. Therefore, the invariants defined for the 2D TSC phase are simply given by 
\begin{align}
	& \nu_{1j} (\VB_{e^{(2)}}, H_{TSC}, H_0) = w_{1j} (\VB_{e^{(2)}}, H_{TSC}), \, j=a, b, c, \nonumber \\
	& \nu_2 (\VB_{e^{(2)}}, H_{TSC}, H_0) = w_2 (\VB_{e^{(2)}}, H_{TSC}), 
\end{align}
and we can simply compute $w_{1j}$ and $w_2$ for the system Hamiltonian $H_{TSC}$ without worrying about the reference Hamiltonian $H_0$.  

We first focus on the regime with $0<m<2$ (denoted as TSC$^+$). 
For the 1D invariants $w_{1j}$, we transform the wavefunctions $\phi_1(\kvec)$ and $\phi_2(\kvec)$ in Eq.~(\ref{eq:TSC_real}) from the real gauge to a smooth gauge along a 1-cell loop $\tilde{l}_j$, while satisfying the $\TR$-constant gauge condition. 
Although the $\TR$-constant and smooth gauge conditions are not achievable over the full BZ for this model (i.e. there is obstruction), we can satisfy such conditions along a single 1-cell loop $\tilde{l}_j$ at a price of having discontinuities away from the loop. 
For the loop $\tilde{l}_a$, since the real-gauge wavefunctions in Eq.~(\ref{eq:TSC_real}) suffer a discontinuity at $k_x=0$, such a gauge transformation to the smooth gauge can only be done by a momentum-dependent transformation matrix
\begin{align}
U(\kvec)=\Theta(k_x)e^{-i\frac{k_x}{2}} \iden_{2\times2}, 
\label{eq:Uk-2d}
\end{align}
where $\iden_{2\times2}$ is in the basis of $\phi_1$ and $\phi_2$, and  $\Theta(k_x)$ is defined as $+1$ for $k_x \geq 0$ and $-1$ for $k_x < 0$. 
Due to the $k_x$-dependent phase factor $e^{-i\frac{k_x}{2}}$ in $U(\kvec)$, the corresponding sewing matrix $\sew_{I/II}=UU^T$ exhibits a $2\pi$-phase winding along the loop $\tilde{l}_a$. We thus find $w_{1a}=1$ using Eq.~(\ref{eq:w1}). 
In contrast, since real-gauge wavefunctions are already smooth along the 1-cell loops $\tilde{l}_b$ and $\tilde{l}_c$, the corresponding sewing matrices exhibit no winding such that we find $w_{1b}=w_{1c}=0$. 

For the 2D invariant, however, we can calculate $w_2 (\VB_{e^{(2)}}, H_{TSC})$ using neither our expression in Sec.~\ref{sec:2d} nor Eq.~(\ref{eq:sum}) since there is obstruction to achieving $\TR$-constant and smooth gauge for the wavefunctions over the full BZ. 
We therefore characterize the 2D TSC phase with helical Majorana  edge modes by 
\begin{align}
    &\nu_{1a}=1,~\nu_{1b}=\nu_{1c}=0\nonumber\\
    &\nu_2: \text{ill-defined}. 
    \label{eq:TSC1_nu}
\end{align}
From a similar calculation, we find that the other regime  $-2<m<0$ (denoted as TSC$^-$) with helical Majorana edge modes is characterized by 
\begin{align}
    &\nu_{1a}=0,~\nu_{1b}=\nu_{1c}=1\nonumber\\
    &\nu_2: \text{ill-defined}. 
    \label{eq:TSC2_nu}
\end{align}
The 2D invariant $\nu_2$ is ill-defined in both regimes due to the same reason, that is the obstruction to satisfying the $\TR$-constant and smooth gauge for the occupied wavefunctions over the full BZ. 
In fact, this obstruction is similar to the obstruction to the $\TR$-constant and smooth gauge in the 2D topological insulator~\cite{Fu2006,Teo2010}. 
Nonetheless, it is desired to detect such a phase by well-defined invariants rather than by the fact that the 2D invariant $\nu_2$ is ill-defined.  

Before ending this subsection, we therefore show that such obstruction can be detected by the well-defined 1D invariants from the condition $\nu_{1a}+\nu_{1c}=1$. 
The real-gauge wavefunctions $\phi_1(\kvec)$ and $\phi_2(\kvec)$ in fact already satisfy the smooth and $\TR$-constant conditions everywhere except at one point, which we refer to as the singular point of the wavefunctions. When we approach the singular point from different directions, the wavefunctions take different values and are thus not well defined at the singular point. This happens because we only take one of the time-reversal-related partners, whereas the well-defined quantity throughout the entire BZ is the 2D Hilbert space spanned by both partners.  
In fact, in both $0<m<2$ and the $-2<m<0$ regimes, there is no gauge in which we can remove this singularity and still satisfy the $\TR$-constant and smooth conditions elsewhere in the BZ. 
However, as we demonstrated when computing $\nu_{1a}$, when we consider a single 1-cell loop that passes through the singular point, we can always remove the singularity while satisfying the smooth condition by a momentum-dependent gauge transformation with phase winding [e.g. the transformation matrix in Eq.~(\ref{eq:Uk-2d})]. 
This will always lead to a nontrivial 1D invariant $\nu_{1j}=1$ along the loop $\tilde{l}_j$ that passes through the singularity. 
In contrast, for a 1-cell loop $\tilde{l}_{j^\prime}$ that does not pass through the singularity, the transformation matrix $U$ from the real gauge to the smooth gauge, and thus the sewing matrix $\sew_{I/II}=UU^T$, do not pick up a phase winding along the loop. 
We therefore expect trivial 1D invariants $\nu_{1j^\prime}=0$ along such loops. 
Since the singularity always occur at either $\tilde{l}_a$ or $\tilde{l}_c$, we conclude that the obstruction to the smooth gauge can be detected by $\nu_{1a}+\nu_{1c}=1$. 
In fact, in our calculations for the $0<m<2$ and the $-2<m<0$ regimes, we find that the singularities occur at $\Gamma$ and $M$, respectively, such that it is the loop $\tilde{l}_a$ and $\tilde{l}_c$ that pass through the singularity, respectively. This explains the 1D invariants we find for the two regimes in Eqs.~(\ref{eq:TSC1_nu}) and (\ref{eq:TSC2_nu}). 

\begin{table*}
\begin{center}
\begin{tabular}{c c | c | c | c | c | c}
	\hline
	Real-space construction & & \thead{Boundary signatures and \\ protecting symmetries} & \hspace{1mm} $\nu_{1a}$ \hspace{1mm} & \hspace{1mm} $\nu_{1b}$ \hspace{1mm} & \hspace{1mm} $\nu_{1c}$ \hspace{1mm} & \hspace{1mm} $\nu_2$ \hspace{1mm} \\ \hline
	Stacking spinful Kitaev chains on $e^{(1)}_1$ &
	\begin{tikzpicture}
		\draw (0,0) [brown,line width=2.5pt,-] --(1,0);
		\draw (0,0) [black,-] --(0,1);
		\draw (0,1) [black,-] --(1,1);
		\draw (1,0) [black,-] --(1,1);
	\end{tikzpicture} & \makecell{Majorana bands along $y$-directional edges;\\ First-order mixed phase}
	& 1 & 0 & 1 & 0 \\ \hline
	Stacking spinful Kitaev chains on $e^{(1)}_3$ &
	\begin{tikzpicture}
		\draw (0,0) [black,-] --(1,0);
		\draw (0,0) [black,-] --(0,1);
		\draw (0,1) [black,-] --(1,1);
		\draw (1,0) [black,-] --(1,1);
		\draw (0,0.5) [brown,line width=2.5pt,-] --(1,0.5);
	\end{tikzpicture} & \makecell{Majorana bands along $y$-directional edges;\\ First-order weak phase}
	& 1 & 0 & 1 & 1 \\ \hline
	Stacking spinful Kitaev chains on $e^{(1)}_2$ &
	\begin{tikzpicture}
		\draw (0,0) [black,-] --(1,0);
		\draw (0,0) [brown,line width=2.5pt,-] --(0,1);
		\draw (0,1) [black,-] --(1,1);
		\draw (1,0) [black,-] --(1,1);
	\end{tikzpicture} & \makecell{Majorana bands along $x$-directional edges;\\ First-order mixed phase}
	& 0 & 1 & 0 & 0 \\ \hline
	Stacking spinful Kitaev chains along $x=\frac{1}{2}$ &
	\begin{tikzpicture}
		\draw (0,0) [black,-] --(1,0);
		\draw (0,0) [black,-] --(0,1);
		\draw (0,1) [black,-] --(1,1);
		\draw (1,0) [black,-] --(1,1);
		\draw (0.5,0) [brown,line width=2.5pt,-] --(0.5,1);
	\end{tikzpicture} & \makecell{Majorana bands along $x$-directional edges;\\ First-order weak phase}
	& 0 & 1 & 0 & 1 \\ \hline
	Stacking TSC$^+$ ($0<m<2$) on $e^{(2)}$ &
	\begin{tikzpicture}
		\draw (0,0) [black,-] --(1,0);
		\draw (0,0) [black,-] --(0,1);
		\draw (0,1) [black,-] --(1,1);
		\draw (1,0) [black,-] --(1,1);
		\draw[fill=blue!15] (0,0) -- (0,1) -- (1,1) -- (1,0) -- cycle;
		\node at (0.5,0.5) {$+$};
	\end{tikzpicture} & \makecell{Helical Majorana edge modes;\\ First-order strong phase}
	& 1 & 0 & 0 & N/A \\ \hline
	Stacking TSC$^-$ ($-2<m<0$) on $e^{(2)}$ &
	\begin{tikzpicture}
		\draw (0,0) [black,-] --(1,0);
		\draw (0,0) [black,-] --(0,1);
		\draw (0,1) [black,-] --(1,1);
		\draw (1,0) [black,-] --(1,1);
		\draw[fill=orange!15] (0,0) -- (0,1) -- (1,1) -- (1,0) -- cycle;
		\node at (0.5,0.5) {$-$};
	\end{tikzpicture} & \makecell{Helical Majorana edge modes;\\ First-order mixed phase}
	& 0 & 1 & 1 & N/A \\ \hline
	Stacking spinful Kitaev chains on $e^{(1)}_1$ and $e^{(1)}_3$ &
	\begin{tikzpicture}
		\draw (0,0) [brown,line width=2.5pt,-] --(1,0);
		\draw (0,0) [black,-] --(0,1);
		\draw (0,1) [black,-] --(1,1);
		\draw (1,0) [black,-] --(1,1);
		\draw (0,0.5) [brown,line width=2.5pt,-] --(1,0.5);
	\end{tikzpicture} & \makecell{Majorana Kramers pairs\\ trapped at two $C_2$-related corners;\\ Second-order strong phase}
	& 0 & 0 & 0 & 1 \\ \hline
\end{tabular}
\end{center}
\caption{A summary of the computed 1D invariants $\nu_{1j}$, $j=a,b,c$ and 2D invariant $\nu_2$ for the representative building-block states of a few important superconducting phases. The bulk-boundary correspondence can be readily established from the results for these states.  
The first column specifies the real-space construction for each of the building-block states. For the schematics on the right, the square denotes a real-space unit cell, a thick brown line denotes a 1D block (spinful Kitaev chain), and a filled square with `$+$' or `$-$' denotes a 2D block of TSC$^{+}$ or TSC$^{-}$ state, respectively. 
The second column specifies the Majorana boundary modes and the protecting symmetries of each phase the building-block state belongs to, and the third to sixth columns are the computed invariants.  
Note that the state in the fourth row is adiabatically connected to the direct sum of the states in the first three rows. The 2D-block states in the fifth and sixth rows, formed by stacking TSC$^+$ and TSC$^-$ states, respectively, are off by a weak phase with Majorana bands on partial edges. However, it is in principle impossible to determine which belongs to the strong phase and which the mixed phase. We choose one to be the strong state and the other is thus fixed as the mixed state. Moreover, since the Majorana bands from the additional weak phase are expected to be buried by the Majorana edges from the strong phase, the two phases are practically speaking indistinguishable.}
\label{tab:building_blocks}
\end{table*}

\section{Diagnosis of boundary modes}
\label{sec:edge}

In this section, we summarize the computed topological invariants, Majorana boundary modes, and protecting symmetries we find for the important building-block states we studied in Sec.~\ref{sec:real} (see Table~\ref{tab:building_blocks}.).
Since each building-block state is adiabatically connected to other states in the same phase, these results allow us to fully diagnose the Majorana boundary modes for any given 2D time-reversal and $C_2$-symmetric superconducting state from our proposed momentum-space invariants. 
Our central results of this work are summarized in Table~\ref{tab:new_inv}, where the final four $\ztwo$ topological invariants $\{z_{1},z_{2},z_{3},z_{4}\}$ we arrive at are combinations of the original invariants $\{\nu_{1a},\nu_{1b},\nu_{1c},\nu_{2}\}$ we find in the momentum-space analysis. These final invariants each diagnose the presence of one type of fundamental Majorana boundary mode protected by a single symmetry. Any mixed phase or a stacking of multiple strong (weak) phases supporting multiple types of boundary modes can be diagnosed by more than one non-zero invariant. 

In the following, we first describe how to extract the Majorana boundary signatures from the original momentum-space invariants $\{\nu_{1a},\nu_{1b},\nu_{1c},\nu_{2}\}$ for any given time-reversal and $C_2$-rotational symmetric superconductor with translational symmetries in the $x$ and $y$ directions. This is summarized in the Fig.~\ref{fig:flow}, and is deduced from the resulting invariants we calculated for each of the building-block states (see Table~\ref{tab:building_blocks}). 
First, we can detect the presence of the first-order strong phase by inequivalent 1D invariants $\nu_{1a}\neq\nu_{1c}$ or an ill-defined 2D invariant $\nu_2$, which results from the obstruction to the $\TR$-constant and smooth conditions (see discussion at the end of Sec.~\ref{sec:TSC}.). Note that this first-order strong phase can also be mixed with any other phases. 
For the rest of the cases with $\nu_{1a}=\nu_{1c}$, the 2D invariant $\nu_2$ is well-defined and can be calculated by either the transition function method (see Sec.~\ref{sec:2d}) or under the interpretation of the second S-W class (see Sec.~\ref{sec:SW}). 
When all the invariants vanish, the state is trivial and has no boundary Majoranas. 
When $\nu_2=1$ is the only nontrivial invariant, the state is in the second-order strong phase with a pair of $C_2$-protected 0D Majorana Kramers pairs. These two pairs of Majorana Kramers pairs are expected to be trapped at $C_2$-related corners, and could in principle annihilate each other by moving along heavily $C_2$-broken edges. 
When the 1D invariant(s) associated with $x$ or $y$-directional 1-cells are also nontrivial besides $\nu_2=1$, i.e. when $\{\nu_{1a},\nu_{1b}, \nu_{1c}\}=\{1,0,1\}$ or $\{0,1,0\}$, the state is in a purely weak phase supporting translation-protected Majorana bands along the $y$- or $x$-directional edges, respectively.  According to our definition for the boundary description, such a weak phase is also considered as a first-order phase.  
Finally, any other combination of invariants indicates that the state is formed by stacking more than one of the above four states. For instance, a vanishing but well-defined $\nu_2$ together with at least one nontrivial 1D invariant indicate that the state is in a phase formed by stacking at least one weak phase with a second-order strong phase. Such a phase is considered a mixed phase since it is protected by both $C_2$ and translational symmetries, and is considered first-order since the corner Majoranas are likely embedded within the Majorana bands along partial edges.

\begin{figure}
	\tikzstyle{do} = [rectangle, draw, fill=gray!20, node distance=1.7cm, text width=15em, text centered, rounded corners, minimum height=2em]
	\tikzstyle{decide} = [draw, ellipse,fill=orange!20, node distance=1.6cm, minimum height=2em]
	\tikzstyle{bd} = [rectangle, draw, fill=blue!20, node distance=3cm, text width=8em, text centered, rounded corners, minimum height=2em]
	\tikzstyle{inv} = [rectangle, text width=5em, text centered, minimum height=2em]
	\tikzstyle{line} = [draw, -latex']
	\vspace{0.5cm}
	\begin{tikzpicture}[node distance = 3cm, auto]
		\node [do] (start) {Calculate $\nu_{1a}$, $\nu_{1b}$, and $\nu_{1c}$};
		\node [decide, below of=start] (stwk) {$\nu_{1a}=\nu_{1c}$?};
		\node [bd, left of=stwk] (strong) {First-order strong phase up to addition of other phases; helical Majorana edge modes};
		\node [do, below of=stwk] (two) {Calculate $\nu_2$ and compare $[\nu_{1a}, \nu_{1b}; \nu_2]$};
		\node [bd, below of=two] (ho) {Second-order strong phase; Majorana corner modes};
		\node [bd, left of=ho] (wk) {First-order weak phase; Majorana bands on partial edges};
		\node [bd, right of=ho] (mix) {First-order mixed phase; Majorana bands on partial edges};
		\path [line] (start) -- (stwk);
		\path [line] (stwk) -- node {no} (strong);
		\path [line] (stwk) -- node {yes} (two);
		\draw [line] (two) -- node [inv] {$[0,0;1]~~~~~~~~~$} (ho);
		\draw [line] (two) -- node [inv,left] {$[1,0;1]$ or $[0,1;1]$ or $[1,1;1]$} (wk);
		\draw [line] (two) -- node [inv,right] {$[1,0;0]$ or $[0,1;0]$ or $[1,1;0]$} (mix);
	\end{tikzpicture}
	\caption{A flowchart that describes how to diagnose the type of Majorana boundary modes from our $\ztwo$ momentum-space topological invariants for a given 2D time-reversal and $C_2$-symmetric superconductor. Here, $\nu_{1a}\neq\nu_{1c}$ implies either a first-order strong state or the direct sum of a first-order strong state and other states. The mixed phase in the bottom right box is the combination of a second-order strong phase and a first-order weak phase.}
	\label{fig:flow}
\end{figure}
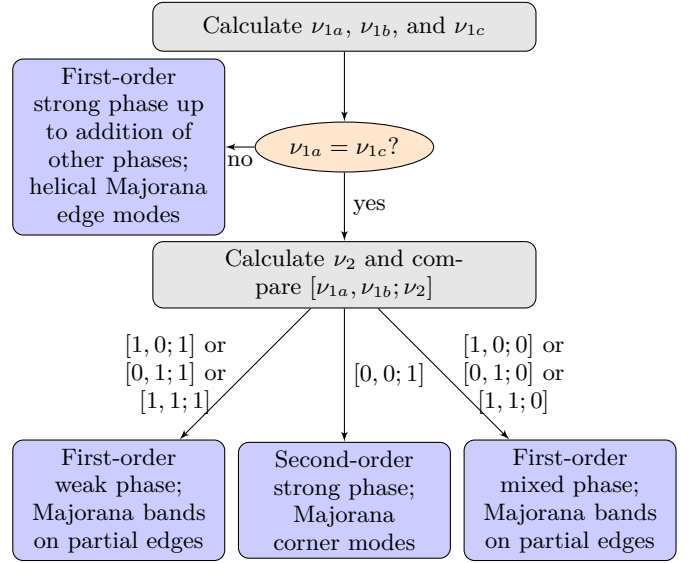

Next, since the 1D and 2D topological invariants $\{\nu_{1a},\nu_{1b},\nu_{1c},\nu_{2}\}$ are not expressed in a basis that naturally has one-to-one correspondence to the type of Majorana boundaries and protecting symmetries, we now explain how to re-express them into a new set of $\ztwo$ invariants $\{z_{1},z_{2},z_{3},z_{4}\}$, which has a simpler relation with the boundary signatures. The results are summarized in Table~\ref{tab:new_inv}. 
Specifically, we start by picking one of the two inequivalent states $TSC^{\pm}$ as the real-space 2D building block. Here, we pick the TSC$^+$ state without loss of generality (the choice made in Table~\ref{tab:building_blocks}) and demand that only one of the new invariants, say $z_1$, is non-vanishing for the resulting first-order strong state built by stacking $TSC^{+}$'s. It is therefore clear that $z_1$ is related to the original invariants by $z_1=\nu_{1a}+\nu_{1c}$. Similarly, by demanding that $z_2$, $z_3$, and $z_4$ are the only nontrivial invariants for a weak phase with Majorana bands along $x$-edges, a weak phase with Majorana bands along $y$-edges, and a second-order strong phase with Majorana corner Kramers pairs, respectively, we arrive at the relation between the new and original invariants in Table~\ref{tab:new_inv}. Note that since $z_4$ depends on $\nu_2$, which is ill-defined in the presence of obstruction to the $\TR$-constant and smooth gauge conditions, $z_4$ can only be computed when $z_1=0$. 
A state formed by stacking more than one of the above four states can be easily detected by more than one nontrivial new invariant $z_i$. In other words, a state with a mixed boundary signature must have more than one non-zero entry in the new set of $\ztwo$ invariants $\{z_{1},z_{2},z_{3},z_{4}\}$. 

It is worth pointing out two subtleties in the implications of these new invariants. 
First, the $C_2$-protected corner Majoranas are only expected to be visible in experiments for a pure second-order strong state characterized by $\{z_{1},z_{2},z_{3},z_{4}\}=\{0,0,0,1\}$. 
When any other invariants are non-zero, the corner Majoranas are expected to be buried by 1D Majorana modes on all or partial edges.
Similarly, Majorana bands on partial edges are expected to be buried by Majorana edge modes on all edges. 
Second, we made an arbitrary choice for the 2D block since it is in principle impossible to determine which of the states built by $TSC^+$ and $TSC^-$ belongs to the pure first-order strong phase. 
Under our choice of $TSC^+$, the other building-block state formed by stacking $TSC^-$ consists of a first-order strong state and some weak state(s). Such a mixed state is characterized by $z_1=1$ along with other non-vanishing invariants $z_2$ and/or $z_3$.  
One can equivalently choose the TSC$^-$ state to be the 2D building block and arrive at \textit{another} self-consistent set of new invariants. 

\begin{table*}
\begin{center}
\begin{tabular}{c | c | c | c}
	\hline
	Redefined invariant & Expression & Boundary signature & Protecting symmetry \\ \hline
	$z_1$ & $\nu_{1a}+\nu_{1c}$ & helical edge modes & time reversal \\ \hline
	$z_2$ & $\nu_{1b}$ & Majorana band & translation along $x$ \\ \hline
	$z_3$ & $\nu_{1c}$ & Majorana band & translation along $y$\\ \hline
	$z_4$ (when $z_1=0$) & $\nu_2+\nu_{1b}\nu_{1c}+\nu_{1b}+\nu_{1c}$ & Majorana corner modes & $C_2$ rotation \\ \hline
\end{tabular}
\end{center}
\caption{A summary of the final set of $\ztwo$ momentum-space invariants that can diagnose the Majorana boundary signatures.  
For each of the new invariants $z_i$, we specify the relation to the original invariants $\nu_{1j}$ and $\nu_2$ (listed in the second column). Each new invariant detects only one type of Majorana boundary mode protected by a single symmetry, as specified in the third and fourth columns.
Any superconducting state that is formed by stacking more than one of these four fundamental states can be detected by a sum of their invariants. 
Since the invariants are $\ztwo$ numbers, the addition and multiplication in the expressions are up to modulo $2$. 
This table is obtained by choosing the 2D-block state built with stacking TSC$^+$ as the pure first-order strong phase with helical Majorana edge modes, which has $\{z_1,z_2,z_3,z_4\}=\{1,0,0,0\}$. By making the other equivalent choice of TSC$^-$, one can arrive at another set of self-consistent new invariants $\tilde{z}_i$.}
\label{tab:new_inv}
\end{table*}

\section{Conclusion and discussion}
\label{sec:discussion}

In this work, we study the classification and the topological invariants for 2D time-reversal superconductors with two-fold rotational symmetry $C_2$. We choose superconductors with this symmetry group because this is one of the simple examples where the symmetry indicators cannot distinguish crystalline topological phases or infer their Majorana boundary features, as observed in previous studies~\cite{Vu2020}. We stress that the $C_2$ symmetry obeying Eq.~(\ref{eq:sym}) is different from the \textit{inversion} symmetry, which squares to $1$ and for which symmetry indicators arise from the parity eigenvalues~\cite{Huang2021}. By establishing the bulk-boundary correspondence for this symmetry class of superconductors using a three-step protocol that some of us previously developed~\cite{Huang2021}, we find topologically distinct phases supporting various Majorana boundary modes, including the first-order phases with edge Majoranas and a second-order phase with $C_2$-protected corner Majoranas.  
Importantly, we show that the topological invariants that can fully distinguish the bulk topology of these phases should depend on band structure information on \textit{high-symmetry lines and general points in the Brillouin zone}, instead of the high-symmetry points. 
Based on our classification results, we derive practically calculable expressions for these invariants, which takes band structures as input and are capable of \textit{diagnosing the Majorana boundary types}.

The three-step protocol we use to derive topological invariants for topological crystalline superconductors consists of the following steps. The first step is to conduct classification study in the momentum space by approximating the K group using the AHSS, where the converged result produces exact subgroups of the K group\cite{Shiozaki2018,Huang2021}.
From our AHSS results, we find that (1) the K group restricted to high-symmetry points is trivial. The topological invariants are therefore not symmetry indicators. (2) The K group restricted to each of the three high-symmetry lines and the K group for general points in the BZ  are both $\ztwo$. The bulk topology can therefore be fully characterized by $\ztwo$ invariants that take band data on high-symmetry lines and general points in the BZ as input. We dub such invariants 1D and 2D invariants, respectively. 
(3) Since the K group contains $\pi_{1}(R_7)$ and $\pi_2(R_7)$ when restricted to 1- and 2-cells, respectively, we write down explicit expressions for the 1D and 2D invariants using the first and second homotopy classes of the sewing matrix for the $C_{2}T$ symmetry, which lives in the space $R_7$.

Based on our AHSS results, we propose practically calculable expressions in the momentum space for three $\ztwo$ 1D invariants defined on the three independent high-symmetry lines (1-cells) and one $\ztwo$ 2D invariant defined on the BZ (2-cells). 
Specifically, we express the 1D invariants $\nu_{1j}$ by the winding number of the sewing matrix along the 1-cell $j$. Importantly, by requiring this winding number to be invariant under gauges that satisfy the $\TR$-constant and smooth conditions, this integer winding number becomes a $\ztwo$ number. 
For the 2D invariant $\nu_2$, we propose the following two expressions for practical computation~\cite{Ahn2018,Ahn2019}. 
The first is obtained by writing the second homotopy class of the sewing matrix defined on the full BZ in terms of the winding number of transition functions between occupied BdG states along non-contractible loops of the BZ. 
Given that this first expression is only practically calculable for specific cases, we also propose a second expression for the 2D invariant. 
Specifically, by identifying the 1D and 2D invariants as the S-W classes of real vector bundles that characterize the system, we can express $\nu_2$ in terms the 1D invariants $\nu_{1j}$'s based on the Whitney sum formula, which we generally know how to calculate.

The second step of our protocol is to conduct a classification study in the real space for the considered class of superconductors. 
The purpose of this step is to understand the Majorana boundary signatures carried by different superconducting phases in this symmetry group, given that the boundary types can be naturally obtained in the real-space picture. 
The method we adopt is the topological crystal approach. This method offers a systematic way to construct representative states (dubbed 'building-block states') for each of the phases, where these building-block states naturally reveal their Majorana boundary types and are adiabatically connected to other states in the same phase. 
From our topological crystal results, we find a $\mathbb{Z}_{2}^{4}$ classification and there are 16 topologically distinct Majorana boundaries, including strong phases with Majorana edges or corners, weak phases with Majorana bands on certain-directional edges, and combinations of these phases.

The third step of our protocol is to establish the bulk-boundary correspondence for the 2D topological crystalline superconductors with time-reversal and $C_2$ symmetries. The purpose of this step is to arrive at topological invariants that do not only discern the bulk topology, but can also predict the type of Majorana boundary modes that we expect to be observed experimentally for a superconductor of interest. 
Given the 1D and 2D topological invariants $\nu_{1j}$'s and $\nu_2$ we obtained in the momentum-space approach and the Majorana boundary types of building-block states we obtained in the real-space approach, we establish the bulk-boundary correspondence by explicitly computing the invariants for each of the building-block states. 

We have summarized our findings from the third step as a flowchart in terms of the $\ztwo$ 1D and 2D invariants $\{\nu_{1a},\nu_{1b},\nu_{1c},\nu_{2}\}$ in Table~\ref{tab:building_blocks} and Fig.~\ref{fig:flow}. Given that these 1D and 2D invariants $\{\nu_{1a},\nu_{1b},\nu_{1c},\nu_{2}\}$ are not expressed in a basis that naturally has one-to-one correspondence to the type of Majorana boundaries and protecting symmetries, we further re-express them into another more intuitive set of invariants $\{z_1,z_2,z_3,z_4\}$. 

This final set of $\ztwo$ topological invariants $\{z_1,z_2,z_3,z_4\}$ is our central result (summarized in Table~\ref{tab:new_inv}). These invariants are \textit{boundary diagnostics} that take band data on the high-symmetry lines and general points in the BZ as inputs. Each invariant diagnoses the presence of one type of fundamental Majorana boundary mode protected by a single symmetry.  
Specifically, $z_1$ diagnoses the presence of the first-order strong phase, which supports time-reversal-protected helical Majorana edge modes. 
Purely weak phases, which support translation-protected Majorana bands along $x$- and $y$-directional edges, are diagnosed by nontrivial $z_2$ and $z_3$, respectively. 
Finally, a nontrivial $z_4$ (in the presence of vanishing $z_1$) diagnoses a stand-alone second-order strong phase, which supports two $C_2$-protected zero-energy Majorana Kramers pairs trapped at $C_2$-related corners.
A combination of the above phases supporting multiple types of boundary modes (possibly embedded in one another) can then be diagnosed by more than one non-zero invariant. 

We make three remarks about our results. First, in Ref.~\cite{Vu2020}, the authors constructed a lattice model for time-reversal and $C_2$-symmetric superconductors. The model realizes four different phases, including a trivial, a higher-order, a weak, and a nodal phase. They argued that there are no symmetry indicators for these systems and numerically showed the boundary signatures including a higher-order phase with Majorana corner modes. In the gapped phases of the lattice model, it is in fact adiabatically connected to the 1D-block states built by stacking of the spinful Kitaev chains, which we have discussed in Sec.~\ref{sec:KC}. We expect our topological invariants still apply even when the Kitaev chains are weakly coupled as long as the gap remains open.  

Second, when a given superconducting state contains a first-order strong state, we expect inevitable obstruction to the $\TR$-constant and smooth gauge conditions, similar to the obstruction in the time-reversal protected 2D topological insulator\cite{Fu2006}. While we can detect such obstruction by the 1D invariant condition $z_1\equiv\nu_{1a}+\nu_{1c}=1$, as we showed in this work, we point out that the 2D invariant $\nu_2$ is ill-defined under such obstruction. In the absence of a first-order strong state, $\nu_2$ is well-defined and can be computed as the second homotopy class of the sewing matrix or as the second S-W class of a vector bundle that characterizes the system, as we propose in this work. In particular, the second S-W class can be obtained from first S-W classes through the Witney sum formula. Other ways of computing the second S-W class have been proposed in previous works for specific cases, such as a method using Wilson loop operators~\cite{Ahn2018,Ahn2019,Li2021}. Another example is when the system is described by a rank-2 vector bundle. In such cases, the second S-W class is equal to the Euler class mod $2$ and has an integral expression as done in Ref. \cite{Ahn2018,Ahn2019, Bouhon2020_1} for insulating systems and semimetals. 

Finally, in this work we have characterized one example of topological crystalline superconducting system, where the AHSS method provides information on the momentum-space topological invariants. We expect that our discussion can extend to other spinful systems with time-reversal and $C_n$ (for an even integer $n$) rotational symmetries. In such systems the combined time-reversal and $C_2$ rotational symmetry is present and the same sewing matrix can be utilized to obtain well-defined topological invariants. While the expressions for the invariants depend on the specific symmetries in the system, the same procedure can be applied to any other symmetry class, where crucial information about the invariants is extracted from the AHSS calculation and the bulk boundary correspondence is established by matching the real-space analysis.

\begin{acknowledgements}

Y. C. thanks Jennifer Cano for helpful discussion. This work was supported by the National Science Foundation under Grant No. PHY 1915165 (T.-C. W. and Y. C.). S.-J. H. acknowledges support from a JQI postdoctoral fellowship and the Laboratory for Physical Sciences.

\end{acknowledgements}

\bibliographystyle{apsrev}	
\bibliography{DIIIwC2}

\begin{thebibliography}{89}
\bibitem[{E1()}]{E1}
\bibinfo{note}{Formally, the definition of the entries in the $E_{1}$ pages are
  given by the relative K group of the pair $(X_{p},X_{p-1})$:
  $E_{1}^{p,-(n+p)} := ^\phi K_G^{(\tau,c),-n}(X_{p},X_{p-1})$, where $X_{p}$
  denotes the $p$-skeleton.}
\bibitem[{Rea()}]{RealGauge}
\bibinfo{note}{Although the real gauge is not exactly the gauge that leaves the
  wavefunctions real everywhere in the BZ, the real gauge is equivalent to the
  latter gauge up to a change of basis and a possible phase factor.}
\bibitem[{bun()}]{bundles}
\bibinfo{note}{In principle a $H^s_{sub}$ can have $4m$ degenerate bands with
  $m\geq2$. In this situation the real vector bundle $\VB_s$ is of rank $2$ or
  more and can have nontrivial second S-W class. We do not expect this to
  happen here because from the real-space classification in Sec.~\ref{sec:real}
  we know that each of the building-block states has a $4$-band Hamiltonian and
  any state is adiabatically connected to a direct sum of the building-block
  states}.
\end{thebibliography}

\appendix

\section{AHSS calculation} \label{app:AHSS}

We follow the procedure introduced in Ref.~\cite{Shiozaki2018} to get the total classification group $^\phi K_G^{(\tau,c),-3}(BZ)$. The calculation is based on the representation interpretation. With the cell decomposition in Fig.~\ref{fig:BZ}, we examine the little group and the emergent AZ class for each independent $p$-cell in a given AZ class $n$. On the $E_1$ page, each entry $E_1^{p,-n}$ is the group classifying the zero-dimensional Hamiltonians of AZ class $n$ with symmetries given by the little group on the $p$-cells. 

\subsection{$0$-cells}
Here we first calculate the emergent AZ class on the $0$-cells. On any of the four 0-cells, the little group is the $C_2$ rotation and the internal symmetries of class $\bar{n}$. There are two irreps $\CR=\pm i$ for the point group $C_2$, related by $\TR$ or $\PH$ if such an anti-unitary symmetry exists. We find that, with class DIII ($\bar{n}=3$), the emergent AZ class for an irrep is AIII because the combined symmetry of $\TR\PH$ maps an irrep back to itself; with class AII ($\bar{n}=4$), the emergent AZ class for an irrep is A since there is no effective symmetry for an irrep; with class CII ($\bar{n}=5$), the emergent AZ class for an irrep is AIII due to the effective symmetry $\TR\PH$. 

\subsection{$1$- and $2$-cells}
On a 1-cell or a 2-cell, the little group is generated by the combined symmetry $\TR\CR$ and/or $\PH\CR$ if $\TR$ and/or $\PH$ exist. In this case the emergent AZ class for class DIII ($\bar{n}=3$) is CI because both $\TR\CR$ and $\PH\CR$ act as effective anti-unitary symmetries where $(\TR\CR)^2=1, (\PH\CR)^2=-1$. The emergent AZ class for class AII ($\bar{n}=4$) is AI due to the effective symmetry $\TR\CR$; the emergent AZ class for class CII ($\bar{n}=5$) is BDI due to the effective symmetries $\TR\CR$ and $\PH\CR$. There are three independent 1-cells and one independent 2-cell, respectively. 

The classification on each representative cell can be read from the zero-dimensional entries for the corresponding emergent AZ classes in the periodic table of topological insulators and superconductors~\cite{Ludwig2015}. The $E_1$ page is shown in Table~\ref{tab:E1}. For example, the diagonal entry $E_1^{0,-3} = \pi_{0}(C_{1}) = 0$, where $C_1 = U(N)$ is the classifying space of the emergent AZ class AIII for $\bar{n} =3$ on the $0$-cells. The entry $E_{1}^{1,-4}$ was obtained from $\pi_{0}(R_{0}) = \mathbb{Z}$ for each 1-cell, where $R_{0} = O(N+M)/(O(N) \times O(M))$ is the classifying space for the emergent AZ class AI for $\bar{n}=4$ on the $1$-cells. The entry $E_{1}^{2,-5}$ was obtained from $\pi_{0}(R_{1}) = \mathbb{Z}_{2}$, where $R_{1} = O(N)$ is the classifying space for the emergent AZ class BDI for $\bar{n} = 5$ on the $2$-cells.

\subsection{First differentials}
We find that the only nontrivial first differential is $d_1^{0,-4}$ since, for the $\bar{n}=4$ row (class AII), the two irreps $\CR=\pm i$ are related by $\TR$ and must coexist on any 0-cell. The generator on the 0-cell thus contributes two states when extended to the adjacent 1-cells. In summary, we have $d_1^{0,-4}$ given by \\
\begin{center}
$d_1^{0,-4}$ =
\begin{tabular}{c | c  c  c  c}
	& $\Gamma$ & $X$ & $Y$ & $M$ \\ \hline
	$a$ & $2$ & $-2$ & $0$ & $0$ \\ \hline
	$b$ & $0$ & $2$ & $0$ & $-2$ \\ \hline
	$c$ & $0$ & $0$ & $2$ & $-2$ \\ 
\end{tabular}
\end{center}
where each entry is the number of generating states formed by an irrep on the 0-cell of the column extended to the 1-cell of the row. Note that the sign is negative if the orientations of the 1-cell and of the adjacent 0-cell disagree. Other first differentials are all trivial. From the first differential and Eq.~(\ref{eq:Er}) we arrive at the $E_2$ page given in Table~\ref{tab:E2}.

The calculation we have done so far is based on the representation interpretation, which serves as a formal calculation tool. It is illuminating to make connection with the topological invariants defined in Sec.~\ref{sec:sew}. Recall that the entry $E_{1}^{1,-4}$ in the $E_{1}$ page was obtained from $\pi_{0}(R_{0})$ for each 1-cell, where $R_{0} = O(N+M)/(O(N) \times O(M))$ is the classifying space for the emergent AZ class AI of the $\bar{n}=4$ row on the 1-cells. Invoking the isomorphism 
\begin{equation}
    \pi_{0}(R_{0}) \cong \pi_{1}(R_{7}),
    \label{eq:isoR0R7}
\end{equation}
where $R_{7} = U(N)/O(N)$, the right-hand side of Eq.~\ref{eq:isoR0R7} corresponds to the 1D topological invariants defined in Sec.~\ref{sec:1d}. We can also understand the first differential $d_{1}^{0,-4} : E_1^{0,-4} \rightarrow E_1^{1,-4}$ as the gauge freedom in the 1D invariants based on the first homotopy group of the sewing matrix as follows. Recall that we obtained the entry $E_{1}^{0,-4}$ from $\pi_{0}(C_{0})$ for each $0$-cell, where $C_{0} = U(N+M)/(U(N) \times U(M))$ is the classifying space for the emergent AZ class A of the $\bar{n}=4$ row on the 0-cells. Moreover, there is an insomorphism
\begin{equation}
    \pi_{0}(C_{0}) \cong \pi_{1}(C_{1}),
    \label{eq:isoC0C1}
\end{equation}
where $C_{1} = U(N)$. Following Ref.~\cite{Ahn2019}, we interpret the gauge freedom of the sewing matrix as the image of the induced map $j^{*}: \pi_{1}(U(N)) \rightarrow \pi_{1}(U(N)/O(N))$ where $j: U(N) \rightarrow U(N)/O(N)$ is the projection. We see that the first differential $d_1^{0,-4}$ consists of such maps for each 1-cell. Therefore, we expect $\text{Im} d_1^{0,-4} = (2\zee)^3$ corresponds to the gauge freedom in the 1D invariants. In Sec.~\ref{sec:1d} we have shown explicitly that gauge freedom indeed leads to the $2\zee$ redundancy on each 1-cell.

\subsection{Second differentials}
The next step is to study the second differentials $d_2^{p,-n}: E_2^{p,-n} \rightarrow E_2^{p+2,-(n+1)}$. We find that only $d_2^{0,-4}$ may be nontrivial. We now turn to the topological phenomena interpretation to calculate the second differential $d_{2}^{0,-4}$.

In the topological phenomena interpretation, a second differential encodes the process where a band inversion occurs at a $0$-cell and generating gapless points in the neighboring $2$-cells. We model this process by the following Hamiltonian around a 0-cell at $\kvec_0$:
\be
	H_{\kvec_0}(\kvec) = [(\kvec-\kvec_0)^2-\mu] \sigma_0\tau_z + (k_x-k_{0,x})\sigma_z\tau_x.
\ee
Time reversal and $C_2$ rotation act as $\TR=i\sigma_y\cc$ and $\CR=i\sigma_z\tau_z$. The gapless points arise when the chemical potential is tuned from $\mu<0$ to $\mu>0$. At one of the gapless points, however, a mass term $(k_y-k_{0,y})\sigma_0\tau_y$ which respects the effective symmetry $\TR\CR$ on the 2-cell can open up a gap. This suggests that $d_2^{0,-4}$, and further $d_2$, is trivial. In a 2D system, $d_r=0$ for any $r>2$. The $E_2$ page is therefore the limiting page $E_\infty$. From the diagonal entries of the limiting page $E_\infty$, we extract the three subgroups of $^\phi K_G^{(\tau,c),-3}(BZ)$:
\begin{align}
	E_{\infty}^{0,-3}=0, \ E_{\infty}^{1,-4}=(\ztwo)^3, \  E_{\infty}^{2,-5}=\ztwo,  
\end{align}

\subsection{Obtaining the K group through the short exact sequences}
In Sec.~\ref{sec:AHSS} we omitted the detailed relation between a generic $^\phi K_G^{(\tau,c),-n}(BZ)$ and its subgroups from the AHSS calculation. We now briefly review it before presenting the result for our system. Denoting the $p$-skeleton in the cell decomposition of the BZ by $X_p$, one can write the classification excluding contributions from subspaces of dimensions lower than $p$ as 
\begin{align}
	F^{p, -(n+p)} = {\rm Ker} \biggl( f: & ^\phi K_G^{(\tau,c),-n}(BZ) \rightarrow \nonumber \\
	& ^\phi K_G^{(\tau,c)|_{X_{p-1}},-n}(X_{p-1}) \biggr),
\end{align}
where $f$ is given by restricting the BZ to its subspace $X_{p-1}$ and $(\tau,c)|_{X_{p-1}}$ describes the symmetry action restricted to $X_{p-1}$. The diagonal entries on the limiting page then satisfy a series of short exact sequences:
\begin{align}
	& 1 \rightarrow F^{1, -(n+1)} \rightarrow \,^\phi K_G^{(\tau, c), -n}(T^d) \rightarrow E_\infty^{0, -n} \rightarrow 1, \nonumber \\
	& 1 \rightarrow F^{2, -(n+2)} \rightarrow F^{1, -(n+1)} \rightarrow E_\infty^{1, -(n+1)} \rightarrow 1, \nonumber \\
	& ... \nonumber \\
	& 1 \rightarrow E_\infty^{d, -(n+d)} \rightarrow F^{d-1, -(n+d-1)} \rightarrow E_\infty^{d-1, -(n+d-1)} \rightarrow 1.
\end{align}
In 2D systems, $^\phi K_G^{(\tau,c),-3}(BZ)$ satisfies:
\begin{align}
	& 1 \rightarrow F^{1,-4} \rightarrow \,^\phi K_G^{(\tau,c),-3}(BZ) \rightarrow E_\infty^{0,-3} \rightarrow 1, \nonumber \\
	& 1 \rightarrow E_\infty^{2,-5} \rightarrow F^{1,-4} \rightarrow E_\infty^{1,-4} \rightarrow 1.
\end{align}
From these and $E_\infty$ in Table~\ref{tab:E2} we reach the result
\be
1 \rightarrow \ztwo \rightarrow ^\phi K_G^{(\tau,c),-3}(BZ) \rightarrow \ztwo^3 \rightarrow 1.
\label{eq:K_ES}
\ee

\section{Stacking two copies of 2D TSC's} \label{app:edge}

We analyze the real-space building blocks by examining their edge signatures. First, we show that the 2D TSC model in Eq.~(\ref{eq:TSC_H}) corresponds to two different phases in the regimes $0<m<2$ and $-2<m<0$. We construct an interface as shown in Fig.~\ref{fig:TSC-TSC} by varying $m$ in the $x$ direction: $0<m(x)<2, \, x>0; \, -2<m(x)<0, \, x<0$. When $m=0$, the gap closes at $(k_x,k_y)=(\pi,0)$ and $(0,\pi)$ in the BZ. The effective Hamiltonian around $(\pi,0)$ is
\be
	H(x,y) = m(x)\tau_z\sigma_0 + i\tau_x\sigma_z\pt_x - i\tau_y\sigma_0\pt_y,
\ee
where $k_x, k_y$ are replaced with $-i\pt_x, -i\pt_y$, respectively. The wavefunction $\Psi(x,y)$ satisfies the eigenequation
\be
	\left[ m(x)\tau_z\sigma_0 + i\tau_x\sigma_z\pt_x \right] \Psi +\left[-E - i\tau_y\sigma_0\pt_y \right] \Psi =0.
\ee
The edge states are of the form $e^{-\int_{x^\prime=0}^x m(x^\prime)dx^\prime} \chi(y)$ where $-\tau_y\sigma_z\chi = \chi$. The two solutions are
\begin{align}
	\chi_1 \propto \frac{1}{\sqrt{2}} \begin{pmatrix} 1 \\ 0 \\ -i \\ 0 \end{pmatrix}, \,
	\chi_2 \propto \frac{1}{\sqrt{2}} \begin{pmatrix} 0 \\ 1 \\ 0 \\ i \end{pmatrix}.
\end{align}
They are related by time reversal and cannot be gapped out. Similarly, the two edge states around $(0,\pi)$ are related to each other by time reversal. The translational symmetry forbids the states around $(\pi,0)$ and those around $(0,\pi)$ to couple and open up a gap. This implies that the two regimes $0<m<2$ and $-2<m<0$ correspond to different phases and differ by a weak phase.

To determine the extension in Eq.~(\ref{eq:K_ES_rs}), we place a 2D TSC next to vacuum, as shown in Fig.~\ref{fig:TSC-vac}. Without loss of generality, we focus on the regime $0<m<2$. The circular geometry is chosen so that we can distinguish the higher-order phase from the trivial phase~\cite{Hsu2020}. The 1D gapped boundary of the former, when $C_2$ symmetry is preserved, will contain gapless points. When $m=2$, the gap closes at $(0,0)$. The effective Hamiltonian around $(0,0)$ is
\be
	H(x,y) = -(2-m(x,y))\tau_z\sigma_0 - i\tau_x\sigma_z\pt_x - i\tau_y\sigma_0\pt_y,
\ee
where $0<m(x,y)<2, \, x^2+y^2<r_0^2; \, m(x,y)>2, \, x^2+y^2>r_0^2$. Changing to polar coordinates, we obtain
\begin{align}
	H(r,\theta) &= -\tilde{m}(r) \tau_z\sigma_0 -i(\cos{\theta}\tau_x\sigma_z + \sin{\theta}\tau_y\sigma_0)\pt_r \nonumber \\
	& + \frac{i}{r} (\sin{\theta}\tau_x\sigma_z - \cos{\theta}\tau_y\sigma_0)\pt_\theta,
	\label{eq:h_circ}
\end{align}
where $\tilde{m} \equiv 2-m$ and $0<\tilde{m}(r)<2, \, r<r_0; \, \tilde{m}(r)<0, \, r>r_0$. The radius of the region $r_0$ should be very large compared to any other scale in the Hamiltonian. The last term in Eq.~(\ref{eq:h_circ}) is much smaller than the other terms and we drop it in the calculation. In this approximation, the wavefunction localized at $r=r_0$ is $\Psi(r,\theta) \approx e^{\int_{r^\prime=r_0}^r \tilde{m}(r^\prime)dr^\prime} \chi(\theta)$, where $(\cos{\theta}\tau_y\sigma_z-\sin{\theta}\tau_x\sigma_0) \chi = \chi$. We write the two solutions as
\begin{align}
	\chi_1 = \frac{\xi_1(\theta)}{\sqrt{2}} \begin{pmatrix}
	e^{-i\theta/2} \\
	0 \\
	ie^{i\theta/2} \\
	0
	\end{pmatrix}, \,
	\chi_2 = \frac{\xi_2(\theta)}{\sqrt{2}} \begin{pmatrix}
	0 \\
	e^{i\theta/2} \\
	0 \\
	-ie^{-i\theta/2}
	\end{pmatrix}.
\end{align}
where $\abs{\xi_{1,2}}=1$. Ref.~\cite{Hsu2020} gives the explicit $\theta$ dependence of $\xi_{1,2}$ in term of the orbital angular momentum quantum number. Now we are equipped to study the direct sum of the 2D TSC to itself. The low-energy Hamiltonian is $H(r,\theta) \otimes \rho_0$, with $\rho$ another Pauli matrix in the space of the two copies. The space of the edge modes is spanned by the following states
\begin{align}
\Psi_1 = e^{\int_{r^\prime=r_0}^r \tilde{m}(r^\prime)dr^\prime} \chi_1(\theta) \otimes \ket{0}, \nonumber \\
\Psi_2 = e^{\int_{r^\prime=r_0}^r \tilde{m}(r^\prime)dr^\prime} \chi_1(\theta) \otimes \ket{1}, \nonumber \\
\Psi_3 = e^{\int_{r^\prime=r_0}^r \tilde{m}(r^\prime)dr^\prime} \chi_2(\theta) \otimes \ket{0}, \nonumber \\
\Psi_4 = e^{\int_{r^\prime=r_0}^r \tilde{m}(r^\prime)dr^\prime} \chi_2(\theta) \otimes \ket{1},
\end{align}
where $\ket{0/1}$ is the $\pm1$ eigenstate of $\rho_z$. $\Psi_1$ and $\Psi_3$ are related by time reversal, and a gap cannot open from the coupling between them; the same is true for $\Psi_2$ and $\Psi_4$. Projecting $H(r,\theta) \otimes \rho_0$ to the edge subspace, we obtain the edge effective Hamiltonian as $-\frac{i}{r} \tau_0\sigma_z\rho_0 \pt_\theta$. A $\kvec$-independent symmetric coupling $\Delta \tau_x\sigma_x\rho_y$ can then open a gap. In the edge subspace, this is
\be
	\Delta \begin{pmatrix}
	0 & 0 & 0 & -\xi_1^*\xi_2 \\
	0 & 0 & \xi_1^*\xi_2 & 0 \\
	0 & \xi_1\xi_2^* & 0 & 0 \\
	-\xi_1\xi_2^* & 0 & 0 & 0.
	\end{pmatrix}
\ee
The gap is therefore non-vanishing everywhere on the circular boundary, in contrast to the boundary of the high-order phase. In conclusion, the direct sum of the 2D TSC to itself is in the trivial phase. To see examples of other symmetry classes where the group extension like the one in Eq.~(\ref{eq:K_ES_rs}) is nontrivial, we refer the readers to e.g. Refs~\cite{Okuma2019,Huang2021}.

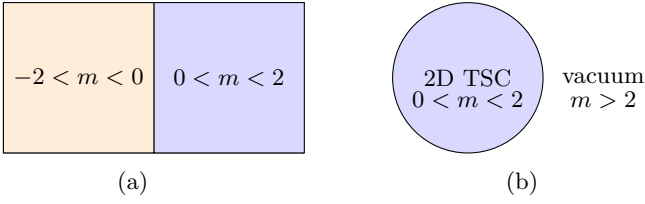
\begin{figure}
	\subcaptionbox{\label{fig:TSC-TSC}}[0.4\linewidth]
		{\begin{tikzpicture}
			\draw (0,0) [black,-] --(2.0,0);
			\draw (0,0) [black,-] --(0,2.0);
			\draw (0,2.0) [black,-] --(2.0,2.0);
			\draw (2.0,0) [black,-] --(2.0,2.0);
			\draw[fill=blue!15] (0,0) -- (0,2.0) -- (2.0,2.0) -- (2.0,0) -- cycle;
			\node at (1,1) {$0<m<2$};
			\draw (0,0) [black,-] --(-2.0,0);
			\draw (0,2.0) [black,-] --(-2.0,2.0);
			\draw (-2.0,0) [black,-] --(-2.0,2.0);
			\draw[fill=orange!15] (0,0) -- (-2.0,0) -- (-2.0,2.0) -- (0,2.0) -- cycle;
			\node at (-1,1) {$-2<m<0$};
		\end{tikzpicture}}
	\hspace{15mm}
	\subcaptionbox{\label{fig:TSC-vac} }[0.4\linewidth]
		{\begin{tikzpicture}
			\tikzstyle{circ} = [draw, shape=circle, fill=blue!15, minimum size=20mm];
			\node[circ] at (0,0) {2D TSC};
			\node at (0,-0.3) {$0<m<2$};
			\node at (1.8,0) {vacuum};
			\node at (1.8,-0.3) {$m>2$};
		\end{tikzpicture}}
	\caption{Configurations to analyze the boundary modes. In (a) two copies of 2D TSC's with different parameter ranges are placed in the half infinite planes next to each other, with a straight edge between them; in (b) there is a 2D TSC within a circle surrounded by vacuum, whose radius is assumed to be large.}
\end{figure}

\section{$\nu_2$ as the winding of the transition function} \label{app:trans_wind}

In the main text we obtain $w_2(\VB_{e^{(1)}_3} \oplus \VB_{e^{(1)}_3}, H_{KC} \oplus H_0)$ by considering the second S-W class. Equivalently, we can calculate it as the winding of the transition function of the real-gauge wavefunctions as described in Sec.~\ref{sec:2d}. In this system $w_{1b}(\VB_{e^{(1)}_3} \oplus \VB_{e^{(1)}_3}, H_{KC} \oplus H_0)=0$, so the winding of $\sew_{I/II}$ in the smooth and $\TR$-constant gauge is trivial along the $k_y$ direction. We pick the loop of $k_x=0/2\pi$ as $l_1$ and the loop of $k_x=\pi/-\pi$ as $l_2$. Then we choose two occupied states that are not related by time reversal and fix them in the real gauge. Using the notation in Eq.~(\ref{eq:KC_real}), one choice is
\begin{align}
	u_1(\kvec) = e^{-i\frac{k_y}{2}} \begin{pmatrix}
	\psi_1(\kvec) \\
	\hline
	0 \\
	0 \\
	0 \\
	0
	\end{pmatrix}, \,
	u_2(\kvec) = e^{-i\frac{k_y}{2}} \begin{pmatrix}
	(0) \\
	\hline
	0 \\
	0 \\
	1 \\
	0
	\end{pmatrix},
\end{align}
where $(0)$ denotes a 4-component zero vector. They obey
\begin{align}
	u_1(-\pi,k_y) = -u_1(\pi,k_y), \nonumber \\
	u_2(-\pi,k_y) = u_2(\pi,k_y), \nonumber \\
	u_1(k_x,-\pi) = -u_1(k_x,\pi), \nonumber \\
	u_2(k_x,-\pi) = -u_2(k_x,\pi).
\end{align}
After bringing all the occupied states to the same energy level, we can carry out a gauge transformation
\begin{align}
	u_1^\prime(\kvec) = \cos(\frac{k_y+\pi}{2}) u_1(\kvec) + \sin(\frac{k_y+\pi}{2}) u_2(\kvec), \nonumber \\
	u_2^\prime(\kvec) = -\sin(\frac{k_y+\pi}{2}) u_1(\kvec) + \cos(\frac{k_y+\pi}{2}) u_2(\kvec),
\end{align}
so that $u_{1,2}^\prime(\kvec)$ still satisfy the real-gauge condition and are smooth along $l_1$. There is a discontinuity across $l_2$, at which the transition function is
\begin{align}
	\braket{u_i^\prime(-\pi,k_y)}{u_j^\prime(\pi,k_y)} = \begin{pmatrix}
	-\cos(k_y+\pi) & \sin(k_y+\pi) \\
	\sin(k_y+\pi) & \cos(k_y+\pi)
	\end{pmatrix}.
\label{eq:t-weak}
\end{align}
The determinant of the transition function Eq.~\ref{eq:t-weak} is $-1$, which signals the nontrivial  $w_{1a}=w_{1c}$. The transition function Eq.~\ref{eq:t-weak} has a winding number $1$ around the loop $l_2$, giving $w_2=1$. Similar calculation can be carried out for the higher-order state to obtain $w_2(\VB_{e^{(1)}_1} \oplus \VB_{e^{(1)}_3} \oplus \VB_{e^{(1)}_3}, H_{KC} \oplus H_{KC} \oplus H_0)$.

\section{Group multiplication} \label{app:group}

The AHSS result itself does not specify the group extension in Eq.~(\ref{eq:K_ES}). Taking the real-space building block states as examples, we can determine the extension by explicitly calculating the group multiplication rule of the K group, i.e. how the invariants add up. Table~\ref{tab:building_blocks} summarizes the invariants for the real-space building block states and some of their direct sums. The group $\ztwo^3$ has trivial action on the group $\ztwo$. We write the group multiplication as
\begin{align}
	[\nu_{1a},\nu_{1b},\nu_{1c};\nu_2] & + [\nu_{1a}^\prime,\nu_{1b}^\prime,\nu_{1c}^\prime;\nu_2^\prime] \nonumber \\
	& = [\nu_{1a}+\nu_{1a}^\prime,\nu_{1b}+\nu_{1b}^\prime,\nu_{1c}+\nu_{1c}^\prime; \nonumber \\
	& \nu_2+\nu_2^\prime+c((\nu_{1a},\nu_{1b},\nu_{1c}), (\nu_{1a}^\prime,\nu_{1b}^\prime,\nu_{1c}^\prime))],
	\label{eq:nu_add}
\end{align}
where $c$ is a 2-cocycle, i.e. $c \in Z^2(\ztwo^3,\ztwo)$. From the group structure in the real-space block sates and the correspondence with the topological invariants in Table~\ref{tab:building_blocks}, we find that there are two possible choices: $c((\nu_{1a},\nu_{1b},\nu_{1c}), (\nu_{1a}^\prime,\nu_{1b}^\prime,\nu_{1c}^\prime))=\nu_{1b}\nu_{1c}^\prime+\nu_{1c}\nu_{1b}^\prime$, or $c((\nu_{1a},\nu_{1b},\nu_{1c}), (\nu_{1a}^\prime,\nu_{1b}^\prime,\nu_{1c}^\prime))=\nu_{1b}\nu_{1a}^\prime+\nu_{1a}\nu_{1b}^\prime$. Since $\nu_{2}$ is ill-defined for a state with $\nu_{1a}\neq\nu_{1c}$, we don't know which one is the correct choice. These two cocycles are in fact 2-coboundaries and therefore correspond to the trivial class in the cohomology group $H^2(\ztwo^3,\ztwo)$. For example, one can remove $c = \nu_{1b}\nu_{1c}^\prime+\nu_{1c}\nu_{1b}^\prime$ in Eq.~(\ref{eq:nu_add}) by redefining the 2D invariant $\nu_2 \rightarrow \nu_2+\nu_{1b}\nu_{1c}$, and similarly for the other choice. This shows that the group extension is trivial and we have  $K=\mathbb{Z}_{2}^{4}$.  However, the group multiplication rule is not fixed due to the ambiguity of the coboundaries.

The group multiplication rule can also be partially deduced from the Whitney sum formula. Without any singular point in the real-gauge wavefunctions, there is a well-defined rank-$N_0$ vector bundle with the entire BZ as its base space. Its first S-W class takes value in the cohomology group $H^1(BZ,\ztwo)$ and its second S-W class in $H^2(BZ,\ztwo)$. In this case $w_{1b}$ and $w_{1c}$ ($=w_{1a}$) correspond to the $\ztwo$ numbers associated with the two non-contractible loops $l(b\cup b^\prime)$ and $l(c\cup c^\prime)$ (in the same homology class as $l(a\cup a^\prime)$), respectively. Similarly, $w_2$ corresponds to the $\ztwo$ number associated with the 2-cycle in the BZ. After taking the reference Hamiltonian into account, we may interpret $\nu_{1j}$'s and $\nu_2$ as the S-W classes of either the original vector bundle or its direct sum with the reference Hamiltonian. The $n=2$ case in Eq.~(\ref{eq:whitney}) suggests that $c$ is determined by the cup product of the first S-W classes for the two summand vector bundles. This is $\nu_{1b}\nu_{1c}^\prime+\nu_{1c}\nu_{1b}^\prime$, or equivalently, $\nu_{1b}\nu_{1a}^\prime+\nu_{1a}\nu_{1b}^\prime$. The choice can be subsequently fixed by the case where a singular point is present and $\nu_{1a}\neq\nu_{1c}$.

\end{document}